\documentclass[12pt]{article}
\usepackage{epsf}
\usepackage{amsmath}
\def\hybrid{\topmargin 0pt      \oddsidemargin 0pt
	\headheight 0pt \headsep 0pt
	\textheight 9in         
	\textwidth 6.25in       
	\marginparwidth .875in
	\parskip 5pt plus 1pt   \jot = 1.5ex}

\catcode`\@=11
\def\marginnote#1{}
\newcount\hour
\newcount\minute
\newtoks\amorpm
\hour=\time\divide\hour by60
\minute=\time{\multiply\hour by60 \global\advance\minute by-\hour}
\edef\standardtime{{\ifnum\hour<12 \global\amorpm={am}%
	\else\global\amorpm={pm}\advance\hour by-12 \fi
	\ifnum\hour=0 \hour=12 \fi
	\number\hour:\ifnum\minute<10 0\fi\number\minute\the\amorpm}}
\edef\militarytime{\number\hour:\ifnum\minute<10 0\fi\number\minute}

\def\draftlabel#1{{\@bsphack\if@filesw {\let\thepage\relax
   \xdef\@gtempa{\write\@auxout{\string
      \newlabel{#1}{{\@currentlabel}{\thepage}}}}}\@gtempa
   \if@nobreak \ifvmode\nobreak\fi\fi\fi\@esphack}
	\gdef\@eqnlabel{#1}}
\def\@eqnlabel{}
\def\@vacuum{}
\def\draftmarginnote#1{\marginpar{\raggedright\scriptsize\tt#1}}

\def\draft{\oddsidemargin -.5truein
	\def\@oddfoot{\sl preliminary draft \hfil
	\rm\thepage\hfil\sl\today\quad\militarytime}
	\let\@evenfoot\@oddfoot \overfullrule 3pt
	\let\label=\draftlabel
\let\marginnote=\draftmarginnote
	\let\marginnote=\draftmarginnote
   \def\@eqnnum{(\theequation)\rlap{\kern\marginparsep\tt\@eqnlabel}%
\global\let\@eqnlabel\@vacuum}  }

\def\numberbysection{\@addtoreset{equation}{section}
\def\theequation{\thesection.\arabic{equation}}}

\def\underline#1{\relax\ifmmode\@@underline#1\else
	$\@@underline{\hbox{#1}}$\relax\fi}

\def\titlepage{\@restonecolfalse\if@twocolumn\@restonecoltrue\onecolumn
     \else \newpage \fi \thispagestyle{empty}\c@page\z@
	\def\thefootnote{\fnsymbol{footnote}} }

\def\endtitlepage{\if@restonecol\twocolumn \else  \fi
	\def\thefootnote{\arabic{footnote}}
	\setcounter{footnote}{0}}  
\catcode`@=12
\relax

\def\beq{\begin{equation}}
\def\eeq{\end{equation}}
\def\bea{\begin{eqnarray}}
\def\eea{\end{eqnarray}}
\def\nn{\nonumber}

\relax
\hybrid

\begin{document}

\begin{titlepage}
\begin{center}
January~2016 \hfill . \\[.5in]
{\large\bf Analytic continuations of 3-point functions of the conformal field theory. }
\\[.5in] 
{\bf Vladimir S.~Dotsenko}\\[.2in]
{\it LPTHE, CNRS, Universit{\'e} Pierre et Marie Curie, Paris VI,\\
               Sorbonne Universit\'es,  75252 Paris Cedex 05, France.}\\[.2in]
               
 \end{center}
 
\underline{Abstract.}

It is shown that the general 3-point function $<\Phi_{c}\Phi_{b}\Phi_{a}>$, with continuous values of charges $a,b,c$ of a statistical model operators, and the 3-point function of the Liouville model, 
could all be obtained by successive analytical continuations starting from the 3-point function 
of the minimal model.

\end{titlepage}

\newpage

\numberwithin{equation}{section}

\section{Introduction.}

Recent interest in the 3-point functions $<\Phi_{c}\Phi_{b}\Phi_{a}>$ with continuous values of charges $a,b,c,$ which do not satisfy the neutrality conditions of the Coulomb gas minimal models, is, 
principally, due to recently found realisations of these correlation functions in the context of statistical models, on the lattice : Potts model 3 spin correlation function [1], loop models [2].

En the other side, the interest in the Liouville model correlation function was always present, since 1981 [3].

The Liouville 3-point function was defined in [4,5] . The statistical model general 3-point function (of imaginary Liouville or Coulomb gas) 
was defined in [6]. See also the related work in [7].

In the present paper we rederive these results somewhat differently, by a sequence of analytical continuations, starting with the minimal model 3-point function [8-10]. 

The present study does not provide new results but gives 
some new methods and provides some unification, hopefully.

\vskip1.5cm

\section{Analytic continuation of $(1,n)$ operators correlation function towards the general $(n',n)$ operators 3-point function.}

The structure constant of the $(1,n)$ minimal model subalgebra, which is the 3-point function of $(1,n)$ operators, is of the form [9]:
\bea
<V^{+}_{1,p}(\infty)V_{1,n}(1)V_{1,m}(0)>\nn\\
=\prod^{k}_{j=1}\frac{\Gamma(j,\rho)}{\Gamma(1-j\rho)}\times\prod^{k-1}_{j=0}\frac{\Gamma(1+\alpha+j\rho)\Gamma(1+\beta+j\rho)\Gamma(1+\gamma+j\rho)}{\Gamma(-\alpha-j\rho)\Gamma(-\beta-j\rho)\Gamma(-\gamma-j\rho)}\label{eq1}
\eea
where $V_{1,m},V_{1,n},V^{+}_{1,p}$, are the Coulomb gas vertex operators,
\bea
V(z,\bar{z})_{1,m}=V_{\alpha_{1,m}}(z,\bar{z})=e^{i\alpha_{1,m}\varphi(z,\bar{z})},\nn\\
V_{1,n}(z,\bar{z})=V_{\alpha_{1,n}}(z,\bar{z})=e^{i\alpha_{1,n}\varphi(z,\bar{z})}\nn\\
\alpha_{1,m}=\frac{1-m}{2}\alpha_{+},\quad\alpha_{1,n}=\frac{1-n}{2}\alpha_{+}\label{eq2}
\eea
$V^{+}_{1,p}$ is the Coulomb gas conjugate operator:
\bea 
V^{+}_{1,p}(z,\bar{z})=V_{\alpha^{+}_{1,p}}(z,\bar{z})=e^{i\alpha^{+}_{1,p}\varphi(z,\bar{z})}\nn\\
\alpha^{+}_{1,p}=2\alpha_{0}-\alpha_{1,p}=2\alpha_{0}-\frac{(1-p)}{2}\alpha_{+}=\alpha_{-}+\frac{1+p}{2}\alpha_{+}\label{eq3}
\eea
$\varphi(z,\bar{z})$ is the Coulomb gas field.

Parameters $\alpha, \beta, \gamma, \rho$ in (\ref{eq1}) are defined as:
\bea
\alpha=2\alpha_{+}\alpha_{1,m}=(1-m)\rho,\quad \beta=2\alpha_{+}\alpha_{1,n}=(1-n)\rho,\nn\\
\gamma=2\alpha_{+}\alpha^{+}_{1,p}=2\alpha_{+}(2\alpha_{0}-\alpha_{1,p})=2\alpha_{+}(\alpha_{-}+\frac{1+p}{2}\alpha_{+})=-2+(1+p)\rho, \quad
\rho=\alpha^{2}_{+}\label{eq4}
\eea
$\alpha_{+},\alpha_{-}$ are the charges of the screening operators
\beq
V_{+}(z,\bar{z})=e^{i\alpha_{+}\varphi(z,\bar{z})},\quad V_{-}(z,\bar{z})e^{i\alpha_{-}\varphi(z,\bar{z})}\label{eq5}
\eeq
$\alpha_{0}$ is the Coulomb gas background charge, $2\alpha_{0}=\alpha_{+}+\alpha_{-}$, and $\alpha_{+}\cdot\alpha_{-}=-1$.

The parameter $k$ in (\ref{eq1}) is the number of screening operators $V_{+}$, required by the function on the l.h.s of (\ref{eq1}), to satisfy the neutrality condition:
\beq
\alpha^{+}_{1,p}+\alpha_{1,n}+\alpha_{1,m}+k\alpha_{+}=2\alpha_{0}\label{eq6}
\eeq
On finds that
\beq
k=\frac{m+n-p-1}{2}\label{eq7}
\eeq

If (\ref{eq1}) is compared with the integral (B.9) in [8], which is the expression for the 3-point function $<V^{+}_{1,p}(\infty)V_{1,n}(1)V_{1,m}(0)>$ with the parameters 
$\alpha$, $\beta$, $\gamma$ given in (\ref{eq4}), 
it is observed that we have removed, in (\ref{eq1}), the factor
\beq
\pi^{k}(\frac{\Gamma(1-\rho)}{\Gamma(\rho)})^{k}\label{eq8}
\eeq
which the normalisation factor. It could be removed by renormalizing the screening operator constant $\mu_{+}$, in the Coulomb gas action for the field $\varphi(z,\bar{z})$. Its general form is
\beq
A[\varphi]=\int d^{2}x(\frac{1}{4\pi}\partial_{z}\varphi\cdot\partial_{\bar{z}}\varphi-\mu_{+}V_{+}-\mu_{-}V_{-})\label{eq9}
\eeq
Also, the last product in (B.9), [9], has been expressed slightly differently: multiplying (\ref{eq6}) by $2\alpha_{+}$ one gets
\bea
\gamma+\beta+\alpha+2\rho k=2\rho-2,\nn\\
\gamma=-2-\alpha-\beta-2\rho(k-1)\label{eq10}
\eea
and the last product in (\ref{eq1}) takes the form
\bea
\prod^{k-1}_{j=0}\frac{\Gamma(1+\gamma+j\rho)}{\Gamma(-\gamma-j\rho)}=\prod^{k-1}_{j=0}\frac{\Gamma(-1-\alpha-\beta-(2k-2-j)\rho)}{\Gamma(2+\alpha+\beta+(2k-2-j\rho)}\nn\\
=\prod^{k-1}_{j=0}\frac{\Gamma(-1-\alpha-\beta-(k-1+j)\rho)}{\Gamma(2+\alpha+\beta+(k-1+j)\rho)}\label{eq11}
\eea
which agrees with (B.9), [9].

The general structure constant of the minimal model, which is the 3-point function of the general $(n',n)$ operators, is of the form [8]:
\bea
<V^{+}_{p',p}(\infty)V_{n',n}(1)V_{m',m}(0)>\nn\\
=\rho^{-4lk}\prod^{l}_{i=1}\frac{\Gamma(i\rho'-k)}{\Gamma(1-i\rho'+k)}\prod^{k}_{j=1}\frac{\Gamma(j\rho)}{\Gamma(1-j\rho)}\nn\\
\times \prod^{l-1}_{i=0}\frac{\Gamma(1-k+\alpha'+i\rho')\Gamma(1-k+\beta'+i\rho')\gamma(1-k+\gamma'+i\rho')}{\Gamma(k-\alpha'-i\rho')\Gamma(k-\beta'-i\rho')\Gamma(k-\gamma'-i\rho')}\nn\\
\times\prod^{k-1}_{j=0}\frac{\Gamma(1+\alpha+j\rho)\Gamma(1+\beta+j\rho)\Gamma(1+\gamma+j\rho)}{\Gamma(-\alpha-j\rho)\Gamma(-\beta-j\rho)\Gamma(-\gamma-j\rho)}\label{eq12}
\eea
Here $V_{m',m},V_{n',n},V^{+}_{p',p}$ are the Coulomb gas vertex operators:
\bea
V_{m',m}(z,\bar{z})=V_{\alpha_{m',m}}(z,\bar{z})=e^{i\alpha_{m',m}\varphi(z,\bar{z})},\nn\\
V_{n',n}(z,\bar{z})=V_{\alpha_{n',n}}(z,\bar{z})=e^{i\alpha_{n',n}\varphi(z,\bar{z})},\nn\\
V^{+}_{p',p}(z,\bar{z})=e^{i\alpha^{+}_{p',p}\varphi(z,\bar{z})}\label{eq13}
\eea
with
\bea
\alpha_{m',m}=\frac{1-m'}{2}\alpha_{-}+\frac{1-m}{2}\alpha_{+},\quad \alpha_{n',n}=\frac{1-n'}{2}\alpha_{-}+\frac{1-n}{2}\alpha_{+},\nn\\
\alpha^{+}_{p',p}=2\alpha_{0}-\alpha_{p',p}=\frac{1+p'}{2}\alpha_{-}+\frac{1+p}{2}\alpha_{+}\label{eq14}
\eea
The parameters $\alpha,\beta,\gamma$ in (\ref{eq12}) are now different from those in (\ref{eq1}), (\ref{eq4}). They are given by:
\bea
\alpha=2\alpha_{+}\alpha_{m',m}=-(1-m')+(1-m)\rho,\quad\beta=2\alpha_{+}\alpha_{n',n}=-(1-n')+(1-n)\rho\nn\\
\gamma=2\alpha_{+}\alpha^{+}_{p',p}=-(1+p')+(1+p)\rho,\quad \rho=\alpha^{2}_{+}\label{eq15}
\eea
and
\bea
\alpha'=2\alpha_{-}\alpha_{m',m}=(1-m')\rho'-(1-m),\quad \beta'=2\alpha_{-}\alpha_{n',n}=(1-n')\rho'-(1-n)\nn\\
\gamma'=2\alpha_{-}\alpha^{+}_{p',p}=(1+p')\rho'-(1+p),\quad \rho'=\alpha^{2}_{-}\label{eq16}
\eea
It is seen that $(\alpha_{+}\alpha_{-}=-1, \,\,
\rho'\alpha_{+}=\alpha_{-}^{2}\alpha_{+}=-\alpha_{-})$:
\bea
\alpha'=-\rho'\alpha,\quad \beta'=-\rho'\beta,\quad\gamma'=-\rho'\gamma,\nn\\
\alpha=-\rho\alpha',\quad \beta=-\rho\beta',\quad \gamma=-\rho\gamma'\label{eq17}
\eea

The parameters $l, k$ in (\ref{eq12}), the numbers of screening operators, they satisfy the Coulomb gas neutrality condition:
\beq
\alpha^{+}_{p',p}+\alpha_{n',n}+\alpha_{m',n}+l\alpha_{-}+k\alpha_{+}=2\alpha_{0}\label{eq18}
\eeq
By collecting the coefficients of $\alpha_{+}$ and $\alpha_{-}$, separately, assuming that there is no compensation between the two ($\rho=\alpha^{2}_{+}$ and $\rho'=\alpha^{2}_{-}=\rho^{-1}$ are being non-rational), one finds:
\beq
l=\frac{m'+n'-p'-1}{2},\quad k=\frac{m+n-p-1}{2}\label{eq19}
\eeq
As compared to the integral (B.10) of [9], which is the expression for the 3-point function 
$<V_{p',p}(\infty)V_{n',n}(1)V_{m',m}(0)>$, we have removed in (\ref{eq12}) 
the normalization factors
\beq
\pi^{l+k}\times(\frac{\Gamma(1-\rho')}{\Gamma(\rho')})^{l}(\frac{\Gamma(1-\rho)}{\Gamma(\rho)})^{k}\label{eq19a}
\eeq
Again, these factors could be removed by renormalising the constants $\mu_{+}$, 
$\mu_{-}$ in (\ref{eq9}).

We have also reorganised the last two factors in the products over $i$ and over $j$, by using the neutrality condition (\ref{eq18}): multiplying (\ref{eq18}) by $2\alpha_{+}$, or by $2\alpha_{-}$, one gets, respectively,
\beq
\gamma+\beta+\alpha-2l+2k\rho=2\rho-2\label{eq20}
\eeq
\beq
\gamma'+\beta'+\alpha'+2l\rho'-2k=-2+2\rho'\label{eq21}
\eeq
which gives
\beq
\gamma=-2-\alpha-\beta+2l-2(k-1)\rho\label{eq22}
\eeq
\beq
\gamma'=-2-\alpha'-\beta'-2(l-1)\rho'+2k\label{eq23}
\eeq
By manipulating the products in (\ref{eq12}), those with $\gamma$ and $\gamma'$, in a way similar to that in (\ref{eq11}), one gets the agreement of (\ref{eq12}) with the expression 
in (B.10), [9].

The objective of this Section is to show that one gets the general 3-point function (\ref{eq12}),
for the general degenerate operators of the minimal model (the operators producing degenerate representations, saying it properly), by replacing $\alpha,\beta, \gamma$ in (\ref{eq1}), (\ref{eq4})
 by $\alpha,\beta,\gamma$ in (\ref{eq15}) and by continuing (\ref{eq1}) to the fractional value of $k$:
\beq
k\rightarrow k-\rho'l\label{eq24}
\eeq
Saying it shortly: (\ref{eq12}) is obtained by the analytic continuation of (\ref{eq1}).

On general remark is in order.

We are talking in this Section about the correlation functions of degenerate operators, in (\ref{eq1}) and in (\ref{eq12}), with $\alpha,\beta,\gamma$ having special values in (\ref{eq4}) and in (\ref{eq15}), to have the objects which are well defined physically, as minimal model correlation functions. But the demonstration given below implies in fact that the Coulomb gas integral in (B.10), [8], with general values of $\alpha,\beta,\gamma$, is obtained from the integral in (B.9) by the analytic continuation in $k$, by eq.(\ref{eq24}). This is up to the normalisation factors (\ref{eq8}) and (\ref{eq19a}).

More precisely, in (\ref{eq1}) and (\ref{eq12}) $\alpha,\beta,\gamma$ would not be totally general. 
They will still be subjects to one constraint, the neutrality condition: (\ref{eq10}) for (\ref{eq1}), 
with $k$ being integer, and (\ref{eq20}) for (\ref{eq12}), with $l,k$ being integers.
As $k$ moves by (\ref{eq24}), the values of $\alpha, \beta, \gamma$ are being moved also,
from the values satisfying (\ref{eq10}) to the values satisfying (\ref{eq20}), like in the case 
of the degenerate values of the parameters, (\ref{eq4}) and (\ref{eq15}).

\vskip0.5cm

Going back to our correlation functions, we shall continue the logarithm of the expression in (\ref{eq1}), by using the integral representation of the logarithm of $\Gamma$-functions.

Let us define
\beq
g_{k}(\rho)=\prod^{k}_{j=1}\frac{\Gamma(j\rho)}{\Gamma(1-j\rho)}\label{eq25}
\eeq
\beq
G_{lk}(\rho)=\prod^{l}_{i=1}\frac{\Gamma(i\rho'-k)}{\Gamma(1-i\rho'+k)}\prod^{k}_{j=1}\frac{\Gamma(j\rho)}{\Gamma(1-j\rho)}\label{eq26}
\eeq
\beq
g^{(\alpha)}_{k}(\rho)=\prod^{k-1}_{j=1}\frac{\Gamma(1+\alpha+j\rho)}{\Gamma(-\alpha-j\rho)}\label{eq27}
\eeq
\beq
G^{(\alpha)}_{lk}(\rho)=\prod^{l-1}_{i=0}\frac{\Gamma(1-k+\alpha'+i\rho')}{\Gamma(k-\alpha'-i\rho')}\prod^{k-1}_{j=0}\frac{\Gamma(1+\alpha+j\rho)}{\Gamma(-\alpha-j\rho)}\label{eq28}
\eeq
and similarly for $g^{(\beta)}_{k}(\rho)$, $g_{k}^{(\gamma)}(\rho)$, $G^{(\beta)}_{lk}(\rho)$,
$G^{(\gamma)}_{lk}(\rho)$.

With these notations, the function in (\ref{eq1}), which we shall note as $C^{p}_{n,m}$, takes the form:
\bea
<V_{1,p}(\infty)V_{1,n}(1)V_{1,m}(0)>\equiv C^{p}_{n,m}(\rho)\nn\\
=g_{k}(\rho)g^{(\alpha)}_{k}(\rho)g^{(\beta)}_{k}(\rho)g_{k}^{(\gamma)}(\rho)\label{eq29}
\eea
and the function in (\ref{eq12}), which we shall note as $C^{(p',p)}_{(n',n)(m',m)}$, takes the form:
\bea
<V^{+}_{p',p}(\infty)V_{n',n}(1)V_{m',m}(0)>\equiv C^{(p',p)}_{(n',n)(m',m)}(\rho)\nn\\
=\rho^{-4lk}G_{lk}(\rho)G_{lk}^{(\alpha)}(\rho)G_{lk}^{(\beta)}(\rho)G_{lk}^{(\gamma)}(\rho)\label{eq30}
\eea

It is shown in the Appendix A that $\log g_{k}(\rho)$, 
analytically continued in $k$, $k\rightarrow k-\rho'l$, eq(\ref{eq24}), is given by:
\beq
\log g_{k-p'l}(\rho)=\log G_{lk}(\rho)-\log \rho\cdot(2kl+l-\rho'l-\rho'l^{2})\label{eq31}
\eeq
For the analytic continuation of $\log g^{(\alpha)}_{k}(\rho)$ we obtain, Appendix A:
\beq
\log g^{(\alpha)}_{k-\rho'l}(\rho)=\log G^{(\alpha)}_{lk}(\rho)
-\log \rho\cdot(2kl-l(2\alpha'-\rho'+1)-l^{2}\rho')\label{eq32}
\eeq
and similar expressions for $\log g^{(\beta)}_{k-\rho'l}(\rho)$ and $\log g^{(\gamma)}_{k-p'l}(\rho)$.

Putting them together, by the eq.(\ref{eq29}), we  obtain
\beq
(\log C^{p}_{n,m}(\rho))_{continued,k-\rho'l}
=\log(G_{lk}G_{lk}^{(\alpha)}G_{lk}^{(\beta)}G^{(\gamma)}_{lk})+\log\rho\cdot(-4kl+2l(\rho'-1))\label{eq33}
\eeq
To get the coefficient of $\log\rho$, in its form above, we have used the neutrality condition on the parameters, eq.(\ref{eq21}).

Finally one obtains:
\beq
(C^{p}_{n,m}(\rho))_{continued, k-\rho'l}=C^{(p',p)}_{n',n)(m',m)}(\rho)\times\rho^{2l(\rho'-1)} \label{eq34}
\eeq
$C^{(p',p)}_{(n',n)(m',m)}(\rho)$ has been reconstructed according to its form in (\ref{eq30}).

The factor $\rho^{2l(\rho'-1)}$ is an another normalisation factor, being an exponent linear in $l$, produced this time in the process of analytic continuation .
It could also be "symmetrized", so that $l$ and $k$ would appear en equal footing, by using the neutrality condition (\ref{eq20}) en $l$ and $k$. We shall do it later, because one extra factor of this type is still coming, will be obtained in the next Section.

For the time being we are dealing with correlation functions of Coulomb gas vertex operators, which is simpler.

They have their own nontrivial normalisation which will be specified later. Further down we shall normalise the operators by 1. With that universal normalisation the extra normalisation factors in our analytic continuation formulas, like the one in (\ref{eq34}), will disappear, as we shall see later. But for the time being we shall still stay with vertex operators, like the ones in (\ref{eq2}), (\ref{eq3}), (\ref{eq13}).

\vskip1.5cm

\section{Analytic continuation to the general, unconstrained values of charges of the vertex operators in the 3-point function.}

To start, we shall reinterpret the results of the previous Section in the opposite direction. We shall consider that it has been shown that the general minimal model 3-point correlation function $<V^{+}_{p',p}(\infty)V_{n',n}(1)V_{m',m}(0)\equiv C^{(p',p)}_{(n',n)(m',m)}$ is equal to the function $<V_{1,p}^{+}(\infty)V_{1,n}(1)V_{1,m}(0)>\equiv C^{p}_{n,m}$ analytically continued, eq.(\ref{eq34}) read from right to left:
\beq
C^{(p',p)}_{(n',n)(m',m)}(\rho)=(C^{p}_{n,m}(\rho))_{continued,k-\rho'l}\times\rho^{-2l(\rho'-1)}\label{eq3.1}
\eeq
And we shall continue further, the function $(C^{p}_{n,m}(\rho))_{continued}$ to the values of charges, of the operators in it, $V^{+}_{p',p}=e^{i\alpha^{+}_{p',p}\varphi}$, $V_{n',n}=e^{i\alpha_{n',n}\varphi}$, $V_{m',m}=e^{i\alpha_{m',m}\varphi}$, to the unconstrained, continuous values, $a,b,c$:
\beq
(V^{+}_{p',p})\rightarrow V_{c}=e^{ic\varphi},\quad V_{n'n}\rightarrow V_{b}=e^{ib\varphi},\quad V_{m',m}\rightarrow V_{a}=e^{ia\varphi}\label{eq3.2}
\eeq
We remind that in the process of analytic continuation in the Section 2, which results in the equality (\ref{eq3.1}), $k$ has been replaced with $k-\rho'l$, but also the charges $\alpha_{1,m}$, $\alpha_{1,n}$, $\alpha_{1,p}^{+}$ has been replaced by $\alpha_{m',m}$, $\alpha_{n',n}$, $\alpha^{+}_{p',p}$, 
comp. the comment preceding the eq.(\ref{eq24}).

It is much easier to continue this way, in two steps, the general minimal model correlator
 $C^{(p',p)}_{(n',n)(m',m)} $,
 \beq
 C^{(p',p)}_{(n',n)(m',m)}\rightarrow (C^{p}_{n,m})_{continued}\rightarrow C^{c}_{b,a}\equiv<V_{c}(\infty)V_{b}(1)V_{a}(0)>\label{eq3.3}
 \eeq
 instead of performing the continuation directly
 \beq
 <V^{+}_{p',p}(\infty)V_{n',n}(1)V_{m',m}(0)>\rightarrow<V_{c}(\infty)V_{b}(1)V_{a}(0)>\label{eq3.4}
 \eeq
  with $a,b,c$, being unconstrained.
 
 To perform the second step in (\ref{eq3.3}), we shall need the detailed expressions for $\log g_{k-\rho'l}(\rho)$, $\log g^{(\alpha)}_{k-\rho'l}(\rho)$, $\log g^{(\beta)}_{k-\rho'l}(\rho)$, $\log g^{(\gamma)}_{k-\rho'l}(\rho)$, obtained in the Appendix A. They are as follows:
 \bea
 \log g_{k-\rho'l}(\rho)=\int_{0}^{\infty}\frac{dt}{t}\{[(k-\rho'l)(k-\rho'l+1)-(k-\rho'l)] e^{-t}\nn\\
 +\frac{(1-e^{-(k-\rho'l)\rho t})e^{-\rho t}+(1-e^{(k-\rho'l)\rho t})e^{-t}}{(1-e^{-t})(1-e^{-\rho t})}\}\label{eq3.5}
 \eea
 \bea
 \log g^{(\alpha)}_{k-\rho'l}(\rho)
 =\int^{\infty}_{0}\frac{dt}{t}\{[(2\alpha+1)(k-\rho'l)+(k-\rho'l-1)(k-\rho'l)\rho] e^{-t}\nn\\
 +\frac{e^{-(1+\alpha)t}(1-e^{-(k-\rho'l)\rho t})
 +e^{\alpha t-\rho t}(1-e^{(k-\rho'l)\rho t})}{(1-e^{-t})(1-e^{-\rho t})}\}\label{eq3.6}
 \eea
 and similar expressions for $\log g^{(\beta)}_{k-\rho l}(\rho)$ and $\log g^{(\gamma)}_{k-p'l}(p)$.
 
For $(C_{n,m}^{p}(p))$ continued we get the following expression:
\bea
\log(C^{p}_{n,m}(\rho))_{continued}=\log g_{k-\rho'l}(\rho)+\log g^{\alpha}_{k-\rho'l}(\rho)+\log g^{(\beta)}_{k-\rho'l}(\rho)+\log g^{(\gamma)}_{k-\rho'l}(\rho)\nn\\
=\int^{\infty}_{0}\frac{dt}{t}\{[(k-\rho'l)(k-\rho'l+1)\rho-(k-\rho'l)\nn\\
+(2\alpha+1+2\beta+1+2\gamma+1)(k-\rho'l)+3(k-\rho'l-1)(k-\rho'l)\rho]e^{-t}\nn\\
+\frac{1}{(1-e^{-t})(1-e^{-\rho t})}[(1-e^{-(k-\rho'l)\rho t})\times(e^{-\rho t}+e^{-(1+\alpha)t}+e^{-(1+\beta)t}+e^{-(1+\gamma)t})\nn\\
+(1-e^{(k-\rho'l)\rho t})(e^{-t}+e^{\alpha t-\rho t}+e^{\beta t-\rho t}+e^{\gamma t-\rho t})]\}\label{eq3.7}
\eea
First we shall simplify the "polynomial" part in (\ref{eq3.7}), the first part of it, the coefficient of $e^{-t}$.

By using the neutrality conditions (\ref{eq20}), on $\alpha$, $\beta$, $\gamma$, we get (we remind that $\rho\rho'=1$):
\bea
(k-\rho'l)(k-\rho'l+1)\rho-(k-\rho'l)\nn\\
+(2\alpha+1+2\beta+1+2\gamma+1)(k-\rho'l)+3(k-\rho'l-1)(k-\rho'l)\rho\nn\\
=(k-\rho'l)[(k-\rho'l+1)\rho-1+2\cdot(2\rho-2+2l-2k\rho)+3 +3(k-\rho'l-1)\rho]\nn\\
=(k-\rho'l)\cdot 2(\rho-1)\label{eq3.8}
\eea
To continue eventually to the general unconstrained values of charges, of the operators, we have to express everything in terms of these charges in particular the combination of the numbers of screenings $k-\rho'l$ in (\ref{eq3.8}).
From (\ref{eq21}):
\beq
2(k-\rho'l)=\alpha'+\beta'+\gamma'+2-2\rho'\label{eq3.9}
\eeq
By eq.(\ref{eq16})
\beq
\alpha'=2\alpha_{-}a,\quad \beta'=2\alpha_{-}b,\quad\gamma'=2\alpha_{-}c\label{eq3.10}
\eeq
where we have replaced $\alpha_{m',m}$, $\alpha_{n',n}$, $\alpha^{+}_{p',p}$ by $a,b,c$. For (\ref{eq3.9}) we obtain
\beq
2(k-\rho'l)=2\alpha_{-}(a+b+c)+2(1-\rho')\label{eq3.11}
\eeq
For (\ref{eq3.8}) we get:
\bea
2(k-\rho'l)(\rho-1)=2\alpha_{-}(a+b+c)(\rho-1)+2(1-\rho')(\rho-1)\nn\\
=-2\alpha_{+}(a+b+c)-2\alpha_{-}(a+b+c)+2(\rho-2+\rho')\nn\\
=-4\alpha_{0}(a+b+c)+8\alpha_{0}^{2}\label{eq3.12}
\eea
We remind that
\beq
\rho=\alpha^{2}_{+},\quad \rho'=\alpha^{2}_{-}\quad\alpha_{+}\alpha_{-}=-1,\quad \rho \alpha_{-}
=-\alpha_{+}\label{eq3.13}
\eeq
For the polynomical part of (\ref{eq3.7}) we obtain:
\beq
\int_{0}^{\infty}\frac{dt}{t}[8\alpha^{2}_{0}-4\alpha_{0}(a+b+c)]e^{-t}\label{eq3.14}
\eeq

We shall simplify next the "exponential" part in (\ref{eq3.7}), its second part. It has to be observed that, separately, the integrals of the polynomial part and of the exponential part in (\ref{eq3.7}), they are divergent at $t\rightarrow 0$. To manipulate them separately we should introduce the limit $\epsilon>0$, in the integrals, instead of 0, and assume that, finally, we shall take the limit $\epsilon\rightarrow 0$
when everything is put together, as it has been done already in the Appendix A.

For the first part we have done no transformations for the integration variable $t$, so we could keep it as it is in (\ref{eq3.14}), although, more properly, we could have assumed that the lower limit of integration in (\ref{eq3.14}) is $\epsilon$.

But for the second part of (\ref{eq3.7}) we do intend to transform the integration variable $t$, so that the explicit introduction of $\epsilon$, for the lower limit of the integration, will be necessary at some point.

In fact, we shall start simplifying (or reorganising) the second part of (\ref{eq3.7}) by transforming the variable $t$:
\beq
t=\sqrt{\rho'}\tilde{t}\label{eq3.15}
\eeq
We shall do all the transformations by ignoring, at first, the divergence at $t=0$. But afterward we shall take specific care of the extra terms,  the "anomaly" terms, being produced by this divergent limit.

With the change of the variable in (\ref{eq3.15}), the second, exponental part of (\ref{eq3.7}) takes form:
\bea
\int^{\infty}_{0}\frac{d\tilde{t}}{d\tilde{t}}\frac{1}{(1-e^{-\sqrt{\rho'}\tilde{t}})(1-e^{-\sqrt{\rho}\tilde{t}})}\nn\\
\times[(1-e^{-(k-\rho'l)\sqrt{\rho}\tilde{t}})(e^{-\sqrt{\rho}\tilde{t}}+e^{-(1+\alpha)\sqrt{\rho'}\tilde{t}}
+e^{-(1+\beta)\sqrt{\rho'}\tilde{t}}+e^{-(1+\gamma)\sqrt{\rho'}\tilde{t}})\nn\\
+(1-e^{(k-\rho'l)\sqrt{\rho}\tilde{t}})(e^{-\sqrt{\rho'}\tilde{t}}+e^{\alpha\sqrt{\rho'}\tilde{t}-\sqrt{\rho}\tilde{t}}+e^{\beta\sqrt{\rho'}\tilde{t}-\sqrt{\rho}\tilde{t}}+e^{\gamma\sqrt{\rho'}\tilde{t}-\sqrt{\rho}\tilde{t}})]\label{eq3.16}
\eea

We shall simplify next the notations and we shall express everything 
$(k-\rho'l$, $\alpha,\beta,\gamma)$ in terms of the charges $a,b,c$.

The Coulomb gas parameter $\sqrt{\rho}=\alpha_{+}$, corresponds (is proportional) 
to the parameter $b$ of the Liouville model, or to the parameter $\beta$ 
in the imaginary Liouville [5,6]. As $b$ and $\beta$ are already in use, 
and the notations $\sqrt{\rho}=\alpha_{+}$, $\sqrt{\rho'}=-\alpha_{-}$ would be slightly heavy, 
we shall use a single parameter, as in the Liouville model, but we shall note it $h$, so that
\beq 
\sqrt{\rho}=\alpha_{+}=h,\quad \sqrt{\rho'}=\frac{1}{h}=-\alpha_{-},\quad \alpha_{0}=\frac{\alpha_{+}+\alpha}{2}=\frac{h}{2}-\frac{1}{2h}\label{eq3.17}
\eeq
Next:
\bea
\alpha=2\alpha_{+}a=2ha,\quad\beta=2hb,\quad\gamma=2hc\nn\\
\alpha'=2\alpha_{-}a=-\frac{2}{h}a\quad\beta'=-\frac{2}{h}b,\quad\gamma'=-\frac{2}{h}c\nn\\
\alpha\sqrt{\rho'}=2a,\quad \beta\sqrt{\rho'}=2b,\quad\gamma\sqrt{\rho'}=2c\nn\\
\alpha'\sqrt{\rho}=-2a,\quad\beta'\sqrt{\rho}=-2b,\quad\gamma'\sqrt{\rho}=-2c\label{eq3.18}
\eea
and, according to (\ref{eq3.9}), or (\ref{eq21}),
\beq
k-\rho'l=\frac{\alpha'+\beta'+\gamma'}{2}+1-\rho'\label{eq3.19}
\eeq
then
\bea
(k-\rho'l)\sqrt{\rho}=\frac{1}{2}(\alpha'+\beta'+\gamma')\sqrt{\rho}+\sqrt{\rho}-\sqrt{\rho'}\nn\\
=-(a+b+c)+\alpha_{+}+\alpha_{-},\nn\\
(k-\rho'l)\sqrt{\rho}=2\alpha_{0}-(a+b+c)\label{eq3.20}
\eea
We shall suppress also the tilde of $\tilde{t}$, $\tilde{t}\rightarrow t$. Then the expression in (\ref{eq3.16}) takes the form:
\bea
\int^{\infty}_{0}\frac{dt}{t}\frac{1}{(1-e^{-\frac{t}{h}})(1-e^{-ht})}\nn\\
\times[(1-e^{-(2\alpha_{0}-a-b-c)t})(e^{-ht}+e^{-\frac{t}{h}-2at}+e^{-\frac{t}{h}-2bt}+e^{-\frac{t}{h}-2ct})\nn\\
+(1-e^{(2\alpha_{0}-a-b-c)t})(e^{-\frac{t}{h}}+e^{-ht+2at}+e^{-ht+2bt}+e^{-ht+2ct})]\label{eq3.21}
\eea
Multiplying the numerator and the denominator in (\ref{eq3.21}) by $\exp\{\frac{t}{2h}+\frac{ht}{2}\}$ we get:
\bea
\int_{0}^{\infty}\frac{dt}{t}\frac{1}{(e^{\frac{t}{2h}}-e^{-\frac{t}{2h}})(e^{\frac{ht}{2}}-e^{-\frac{ht}{2}})}\nn\\
\times[(1-e^{-(2\alpha_{0}-a-b-c)t})(e^{-\alpha_{0}t}+e^{\alpha_{0}t-2at}+e^{\alpha_{0}t-2bt}+e^{\alpha_{0}t-2ct})\nn\\
+(1-e^{(2\alpha_{0}-a-b-c)t})(e^{\alpha_{0}t}+e^{-\alpha_{0}t+2at}+e^{-\alpha_{0}t+2bt}+e^{-\alpha_{0}t+2ct})]\label{eq3.22}
\eea
Next we obtain, by regrouping the terms:
\bea
\int^{\infty}_{0}\frac{dt}{t}\frac{1}{4\sinh\frac{t}{2h}\cdot\sinh\frac{ht}{2}}\nn\\
\times[e^{\alpha_{0}t}+e^{-\alpha_{0}t}-e^{(3\alpha_{0}-a-b-c)t}-e^{-(3\alpha_{0}-a-b-c)t}\nn\\
+e^{(\alpha_{0}-2a)t}+e^{-(\alpha_{0}-2a)t}+e^{(\alpha_{0}-2b)t}+e^{-(\alpha_{0}-2b)t}+e^{(\alpha_{0}-2c)t}+e^{-(\alpha_{0}-2c)t}\nn\\
-e^{(\alpha_{0}+a-b-c)t}-e^{-(\alpha_{0}+a-b-c)t}-e^{(\alpha_{0}-a+b-c)t}-e^{-(\alpha_{0}-a+b-c)t}\nn\\
-e^{(\alpha_{0}-a-b+c)t}-e^{-(\alpha_{0}-a-b+c)t}]\label{eq3.23}
\eea  
It can be presented as:
\bea
\int_{0}^{\infty}\frac{dt}{t}\frac{1}{\sinh\frac{t}{2h}\cdot\sinh\frac{ht}{2}}
\times[\sinh^{2}(\alpha_{0}\frac{t}{2})\nn\\
+\sinh^{2}((\alpha_{0}-2a)\frac{t}{2})
+\sinh^{2}((\alpha_{0}-2b)\frac{t}{2})
+\sinh^{2}((\alpha_{0}-2c)\frac{t}{2})\nn\\
-\sinh^{2}((3\alpha_{0}-a-b-c)\frac{t}{2})\nn\\
-\sinh^{2}((\alpha_{0}+a-b-c)\frac{t}{2})
-\sinh^{2}((\alpha_{0}-a+b-c)\frac{t}{2})
-\sinh^{2}((\alpha_{0}-a-b+c)\frac{t}{2})]\label{eq3.24}
\eea
For $t\rightarrow 0$, the above integral takes the asymptotic form:
\bea
\int_{0}\frac{dt}{t}\frac{4}{t^{2}}\times\frac{t^{2}}{4}[(\alpha_{0})^{2}+(\alpha_{0}-2a)^{2}
+(\alpha_{0}-2b)^{2}+(\alpha_{0}-2c)^{2}-(3\alpha_{0}-a-b-c)^{2}\nn\\
-(\alpha_{0}+a-b-c)^{2}-(\alpha_{0}-a+b-c)^{2}-(\alpha_{0}-a-b+c)^{2}]\nn\\
=\int_{0}\frac{dt}{t}[-8\alpha^{2}_{0}+4\alpha_{0}(a+b+c)]\label{eq3.25}
\eea
This divergence, at $t\rightarrow 0$, is compensated by the polynomial part (\ref{eq3.14}), of the integral(\ref{eq3.7}). 
But, saying it differently, the equality of the expressions under integrals in (\ref{eq3.14}) and (\ref{eq3.25}) implies that 
the polynomial part (\ref{eq3.14}) could be distributed as in (\ref{eq3.25}):
\bea
-\int^{\infty}_{0}\frac{dt}{t}[(\alpha_{0})^{2}+(\alpha_{0}-2a)^{2}+(\alpha_{0}-2b)^{2}+\alpha_{0}-2c)^{2}
-3(\alpha_{0}-a-b-c)^{2}\nn\\
-(\alpha_{0}+a-b-c)^{2}-(\alpha_{0}-a+b-c)^{2}-(\alpha_{0}-a-b+c)^{2}]\times e^{-t}\label{eq3.26}
\eea
so that the full integral (\ref{eq3.7}), which is the sum of (\ref{eq3.14})=(\ref{eq3.26}) and  (\ref{eq3.24}), takes the form:
\bea
\int_{0}^{\infty}\frac{dt}{t}\{-[(\alpha_{0})^{2}e^{-t}
-\frac{\sinh^{2}(\alpha_{0}\frac{t}{2})}{\sinh\frac{t}{2h}\cdot\sinh\frac{ht}{2}}]
-[(\alpha_{0}-2a)^{2}e^{-t}-\frac{\sinh^{2}((\alpha_{0}-2a)\frac{t}{2})}{\sinh\frac{t}{2h}\cdot\sinh\frac{ht}{2}}]\nn\\
-[(\alpha_{0}-2b)^{2}e^{-t}-\frac{\sinh^{2}((\alpha_{0}-2b)\frac{t}{2})}{\sinh\frac{t}{2h}\cdot\sinh\frac{ht}{2}}]
-[(\alpha_{0}-2c)^{2}e^{-t}-\frac{\sinh^{2}((\alpha_{0}-2c)\frac{t}{2})}{\sinh\frac{t}{2h}\cdot\sinh\frac{ht}{2}}]\nn\\
+[(3\alpha_{0}-a-b-c)^{2}e^{-t}-\frac{\sinh^{2}((3a-a-b-c)\frac{t}{2})}{\sinh\frac{t}{2h}\cdot\sinh\frac{ht}{2}}]\nn\\
+[(\alpha_{0}+a-b-c)^{2}e^{-t}-\frac{\sinh^{2}((\alpha_{0}+a-b-c)\frac{t}{2})}{\sinh\frac{t}{2h}\cdot\sinh\frac{ht}{2}}]\nn\\
+[(\alpha_{0}-a+b-c)^{2}e^{-t}-\frac{\sinh^{2}((\alpha_{0}-a+b-c)\frac{t}{2})}{\sinh\frac{t}{2h}\cdot\sinh\frac{ht}{2}}]\nn\\
+[(\alpha_{0}-a-b+c)^{2}e^{-t}-\frac{\sinh^{2}((\alpha_{0}-a-b+c)\frac{t}{2})}{\sinh\frac{t}{2h}\cdot\sinh\frac{ht}{2}}]\}\label{eq3.27}
\eea
We find that everything is expressed in terms of the function $\Upsilon(x,h)$ [5,6]
\beq
\log\Upsilon_{M}(x,h)=\int_{0}^{\infty}\frac{dt}{t}\{(\alpha_{0}-x)^{2}e^{-t}
-\frac{\sinh^{2}((\alpha_{0}-x)\frac{t}{2})}{\sinh\frac{t}{2h}\cdot\sinh\frac{ht}{2}}\}\label{eq3.28}
\eeq
so that $\log(C^{p}_{n,m})_{continued}$, which is the integral (\ref{eq3.7}), takes the form:
\bea
\log(C^{p}_{n,m})_{continued}=-\log\Upsilon_{M}(2\alpha_{0},h)\nn\\
-\log\Upsilon_{M}(2a,h)-\log\Upsilon_{M}(2b,h)-\log\Upsilon_{M}(2c,h)\nn\\
+\log\Upsilon_{M}(a+b+c-2\alpha_{0},h)\nn\\
+\log\Upsilon_{M}(-a+b+c,h)+\log\Upsilon_{M}(a-b+c,h)+\log\Upsilon_{M}(a+b-c,h)\label{eq3.29}
\eea
and
\bea
(C^{p}_{n,m})_{continued}\nn\\
=\frac{\Upsilon_{M}(a+b+c-2\alpha_{0},h)
\Upsilon_{M}(-a+b+c,h)\Upsilon_{M}(a-b+c,h)\Upsilon_{M}(a+b-c,h)}{\Upsilon_{M}(2\alpha_{0},h)
\Upsilon_{M}(2a,h)\Upsilon_{M}(2b,h)\Upsilon_{M}(2c,h)}\label{eq3.30}
\eea
This is the function in [6], though not completely so.

First, the normalisation of operators used in [6] is different.

Second, we have putted the index $"M"$ for $\Upsilon_{M}$, for "matter" (statistical model) which is slightly different from $\Upsilon(x,h)$ for Liouville, gravity, which has been introduced in [5]. The difference is in:
\beq
\alpha_{0}=\frac{h}{2}-\frac{1}{2h}, \,\,\mbox{instead of} \quad b_{0}=\frac{h}{2}
+\frac{1}{2h}, \,\, \mbox{for} \,\,\Upsilon(x,h)  \,\,\mbox{of Liouville}\label{eq3.31}
\eeq
In [6], the function which note $\Upsilon_{M}(x,h)$ has been used in the form $\Upsilon(x+\frac{1}{h},h)$.

And third, there is one additional factor missing in (\ref{eq3.30}), the way we derived it.
We haven't calculated yet the anomaly term, which is produced because of our manipulations with the second, exponential part integral, which is divergent at $t\rightarrow 0$ when taken separately. So far, in our derivation of (\ref{eq3.30}), we have ignored this point.
We shall take care of it now.

The second, exponential part in (\ref{eq3.7}) should have been taken with the lower integration limit $\epsilon$, instead of 0, before the change of the variable $t$ in (\ref{eq3.15}). We reproduce this integral somewhat symbolically:
\beq
\lim_{\epsilon\rightarrow 0}\int^{\infty}_{\epsilon}\frac{dt}{t}\frac{1}{(1-e^{-t})(1-e^{-\rho t})}[t,...]\label{eq3.32}
\eeq
After the change of the variable in (\ref{eq3.15}) we obtain
\beq
\lim_{\epsilon\rightarrow 0}\int^{\infty}_{\frac{\epsilon}{\sqrt{\rho'}}}\frac{d\tilde{t}}{\tilde{t}}\cdot\frac{1}{(1-e^{-\sqrt{\rho'}\tilde{t}})(1-e^{-\sqrt{\rho}\tilde{t}})}[\sqrt{\rho'}\tilde{t},...]\label{eq3.33}
\eeq
The integral can be decomposed as follows:
\bea 
\lim_{\epsilon\rightarrow 0}\int^{\infty}_{\epsilon}\frac{d\tilde{t}}{\tilde{t}}
\cdot\frac{1}{(1-e^{-\sqrt{\rho'}\tilde{t}})(1-e^{-\sqrt{\rho}\tilde{t}})}[\sqrt{\rho'}\tilde{t},...]\nn\\
-\int^{\frac{\epsilon}{\sqrt{\rho'}}}_{\epsilon} \frac{d\tilde{t}}{\tilde{t}}\cdot\frac{1}{(1-e^{-\sqrt{\rho'}\tilde{t}})(1-e^{-\sqrt{\rho}\tilde{t}})}[\sqrt{\rho'}\tilde{t},...]\label{eq3.34}
\eea
The first integral in above goes to join the first, polynomial part of (\ref{eq3.7}) and gives finally the function in (\ref{eq3.30}). But the second integral in (\ref{eq3.34}) gives an additional term, which has been missed in our derivation of (\ref{eq3.30}).

To calculate the second integral in (\ref{eq3.34}) we could use all the transformations, for the expression under the integral, which has been done above. It could be taken in the form in (\ref{eq3.24}), but with the limits of integration ($\epsilon,\epsilon/\sqrt{\rho'}$) instead of $(0,\infty)$. As $t$ stays small, in the limits $(\epsilon,\epsilon/\sqrt{\rho'})$, we can replace the expression under the integral by its limiting form, for $t\rightarrow 0$, which has already been obtained in (\ref{eq3.25}). In this way we get, for the second integral in (\ref{eq3.34}), the following result:
\bea
-\int_{\epsilon}^{\epsilon/\sqrt{\rho'}}\frac{dt}{t}[-8\alpha^{2}_{0}+4\alpha_{0}(a+b+c)]\nn\\
=[8\alpha_{0}^{2}-4\alpha_{0}(a+b+c)]\cdot(\log\frac{\epsilon}{\sqrt{\rho'}}-\log\epsilon)\nn\\
=\log\rho\cdot(4\alpha^{2}_{0}-2\alpha_{0}(a+b+c))\label{eq3.35}
\eea
This is our anomaly. It has to be added to (\ref{eq3.29}). With it, the formula in (\ref{eq3.30}) takes the form:
\bea
(C^{p}_{n,m})_{continued}\nn\\
=\frac{\Upsilon_{M}(a+b+c-2\alpha_{0})\Upsilon_{M}(-a+b+c)\Upsilon_{M}(a-b+c)\Upsilon_{M}(a+b-c)}{\Upsilon_{M}(2\alpha_{0})\Upsilon_{M}(2a)\Upsilon_{M}(2b)\Upsilon_{M}(2c)}\nn\\
\times\rho^{4\alpha^{2}_{0}-2\alpha_{0}(a+b+c)}\label{eq3.36}
\eea

We remind that our objective was to continue the general minimal model 3-point function
\beq
C^{(p',p)}_{(n',n)(m',m)}=<V^{+}_{p',p}(\infty)V_{n',n}(1)V_{m',m}(0)>\label{eq3.37}
\eeq
towards the function
\beq
<V_{c}(\infty)V_{b}(1)V_{a}(0)>\label{eq3.38}
\eeq
with $a,b,c$ unconstraint. By (\ref{eq3.1}), the result of the first step of continuation, and (\ref{eq3.36}), the result of the second step of continuation, we find so far:
\bea
C^{(p',p)}_{(n',n)(m',m)}=<V^{+}_{p',p}(\infty)V_{n',n}(1)V_{m',m}(0)>\nn\\
=\frac{\Upsilon_{M}(a+b+c-2\alpha_{0})\Upsilon_{M}(-a+b+c)\Upsilon_{M}(a-b+c)\Upsilon_{M}(a+b-c)}{\Upsilon_{M}(2a)\Upsilon_{M}(2b)\Upsilon_{M}(2c)}\nn\\
\times\rho^{4\alpha^{2}_{0}-2\alpha_{0}(a+b+c)}\times\rho^{2l(1-\rho')}\label{eq3.39}
\eea

We have suppressed the factor $\Upsilon_{M}(2\alpha_{0})$ in the denominator of (\ref{eq3.36}), because $\Upsilon_{M}(2\alpha_{0})=1$, as can be checked directly using the integral definition of the function $\Upsilon_{M}(x)$ in (\ref{eq3.28}). Some specific values of $\Upsilon_{M}(x)$ are listed in the Appendix B. With respect to the notations, we suppress sometimes the dependence of $\Upsilon_{M}$ on $h$, which is implicite, by writing $\Upsilon_{M}(x)$ for $\Upsilon_{M}(x,h)$, as in (\ref{eq3.36}) and in (\ref{eq3.39}).

In eq.(\ref{eq3.39}), the charges $a,b,c$ in the r.h.s. are still having the discretized values of the degenerate charges:
\beq
a=\alpha_{m',m},\quad b=\alpha_{n',n},\quad c=\alpha^{+}_{p',p}\label{eq3.40}
\eeq
The remaining obstacle to fully continue to the continuous values of $a,b,c$, is the factor
\beq
\rho^{2l(1-\rho')}\label{eq3.41}
\eeq
in (\ref{eq3.39}). In particular, $l$ is still given by (\ref{eq19}).

The total $\rho$-factor in (\ref{eq3.39}) could be symmetrized. It is easy to check, by using the equation (\ref{eq3.20}) for $k-\rho'l$, that
\beq
\rho^{4\alpha^{2}_{0}-2\alpha_{0}(a+b+c)+2l(1-\rho')}=(\rho)^{l(1-\rho')}\times(\rho')^{k(1-\rho)}\label{eq3.42}
\eeq

Still, in the $\rho$-factor in (\ref{eq3.39}), or in (\ref{eq3.42}), there appear the numbers of screenings, 
$k$ and $l$, in the form which could not be expressed fully by the charges $a,b,c$.

But it is clear, by the form of the $\rho$-factor in (\ref{eq3.42}), that it could be removed by the renormalisation of the constants $\mu_{+}$ and $\mu_{-}$ in (\ref{eq9}).

Specifically, if we give the following values for $\mu_{+}$, $\mu_{-}$:
\bea
\mu_{+}=\frac{1}{\pi}\frac{\Gamma(\rho)}{\Gamma(1-\rho)}\times(\rho')^{-(1-\rho)},\nn\\
\mu_{-}=\frac{1}{\pi}\frac{\Gamma(\rho')}{\Gamma(1-\rho')}\times\rho^{-(1-\rho')}\label{eq3.43}
\eea
then the normalisation factor (\ref{eq19a}) will disappear, from the result for the integral (B.10) in [8], and the factor
\beq
\rho^{-l(1-\rho')}\times(\rho')^{-k(1-\rho)}\label{eq3.44}
\eeq
will appear, in front of the expression for $<V^{+}_{p',p}(\infty)V_{n',n}(1)V_{m',m}(0)>$ in (\ref{eq12}). Then we perform the transformations of the Section 2 and of the present Section, to arrive to (\ref{eq3.39}), but because of the extra factor (\ref{eq3.44}) always present, unchanged during our transformations, the extra $\rho$ factor in (\ref{eq3.39}), in its form in (\ref{eq3.42}), will be cancelled. We shall get, with the choice (\ref{eq3.43}) for the values of the Coulomb gas  
constants $\mu_{+}$, $\mu_{-}$, the formula (\ref{eq3.39}), without the $\rho$-factor. At this point we could finally continue $a,b,c$ to the continuous values and we obtain, finally, the formula for 3-point function in the form
\bea
<V_{c}(\infty)V_{b}(1)V_{a}(0)>\nn\\
=\frac{\Upsilon_{M}(a+b+c-2\alpha_{0})\Upsilon_{M}(-a+b+c)\Upsilon_{M}(a-b+c)\Upsilon_{M}(a+b-c)}{\Upsilon_{M}(2a)\Upsilon_{M}(2b)\Upsilon_{M}(2c)} \label{eq3.45}
\eea
with $a,b,c$ taking general, continuous values.

We summarise that the formula (\ref{eq3.45}) has all been obtained by the analytic continuation from general 3-point function for degenerate operators of the minimal model. Though a specific normalisation of the Coulomb gas screening operators, or of the constants $\mu_{+}$, $\mu_{-}$, was required.

The formula (\ref{eq3.45}) is that for the vertex operators
\beq
V_{a}(z,\bar{z})=e^{ia\varphi(z,\bar{z})},\quad V_{b}(z,\bar{z})=e^{ib\varphi(z,\bar{z})},\quad
V_{c}(z,\bar{z})=e^{ic\varphi(z,\bar{z})}\label{eq3.46}
\eeq
with their nontrivial normalisation, $N_{a}$ for $V_{a}(z,\bar{z})$, etc., which will be specified in the next Section.

In the next Section we shall get a slightly different formula, compared to (\ref{eq3.45}), for the 3-point function $<\Phi_{c}(\infty)\Phi_{b}(1)\Phi_{a}(0)>$ of the normalised operators:
\beq
\Phi_{a}(z,\bar{z})=\frac{1}{N_{a}}V_{a}(z,\bar{z}),\quad \Phi_{b}(z,\bar{z})=\frac{1}{N_{b}}V_{b}(z,\bar{z}),\quad
\Phi_{c}(z,\bar{z})=\frac{1}{N_{c}}V_{c}(z,\bar{z})\label{eq3.47}
\eeq
In the case of normalised operators (\ref{eq3.47}) the $\rho$ factors get cancelled automatically, 
independently of the choice of normalisation of the screening operators, in the course of our derivation 
from the original Coulomb gas formulas.

\vskip1.5cm

\section{Normalisations. 3-point function of normalised operators.}

We shall fix the normalisation of Coulomb gas vertex operators, to normalise them finally as in (\ref{eq3.47}), by analysing the values of the correlation functions calculated 
for $a,b,c$ having discrete, degenerate values
\beq
a=\alpha_{m',m},\quad b=\alpha_{n',n},\quad c=\alpha_{p',p}\label{eq4.1}
\eeq
We shall do it by using the original expression for the correlation functions, in terms of products of $\Gamma$ functions, in its symmetrized form given below, and also by using the new expression, in terms of products of $\Upsilon$ functions, which should give the same values, when $a,b,c$ are degenerate, eq.(\ref{eq4.1}).

The fact that we have kept, in Sections 2 and 3, the charge $\alpha_{p',p}$ always in 
its conjugate form
\beq
c=\alpha^{+}_{p',p}=2\alpha_{0}-\alpha_{p',p}\label{eq4.2}
\eeq 
is not actually important for our derivations. We can relax now to the values of $c$ in (\ref{eq4.1}). 
The case of $c=\alpha^{+}_{p',p}$ will correspond, with the definition of $c$ in (\ref{eq4.1}), 
to $\alpha_{-p',-p}$, instead of $\alpha_{p',p}$. Which means that we shall allow 
for the indices to take also the negative values.
We have kept, in the derivations of Section 2 and 3, one of the operators, in its conjugate form, $V^{+}_{p',p}(z,\bar{z})$, in part for historical reasons, 
to make the transition from the original formulas of [9] smoother, not to become excessively general from the start, which is not needed. 

We shall go back, in this Section, to the normalisation of $\mu_{+}$, $\mu_{-}$:
\beq
\mu_{+}=\frac{1}{\pi}\frac{\Gamma(\rho)}{\Gamma(1-\rho)},\quad\mu_{-}=\frac{1}{\pi}\frac{\Gamma(\rho')}{\Gamma(1-\rho')}\label{eq4.3}
\eeq
which has been taken at the start, in the Section 2, with which the correlation function of vertex operators is of the form:
\bea
<V_{c}(\infty)V_{b}(1)V_{a}(0)>=\rho^{-4lk}\prod^{l}_{i=1}\frac{\Gamma(i\rho'-k)}{\Gamma(1-i\rho'+k)}\times\prod^{k}_{j=1}\frac{\Gamma(j\rho)}{\Gamma(1-j\rho)}\nn\\
\times\prod^{l-1}_{i=0}\frac{\Gamma(1-k+\alpha'+i\rho')\Gamma(1-k+\beta'+i\rho')\Gamma(i-k\gamma'+i\rho')}{\Gamma(k-\alpha'-i\rho')\Gamma(k-\beta'-i\rho')\Gamma(k-\gamma'-i\rho')}\nn\\
\times\prod^{k-1}_{j=0}\frac{\Gamma(1+\alpha+j\rho)\Gamma(1+\beta+j\rho)\Gamma(1+\gamma+j\rho)}{\Gamma(-\alpha-j\rho)\Gamma(-\beta-j\rho)\Gamma(-\gamma-j\rho)}\label{eq4.4}
\eea
when $a,b,c$ take the degenerate values (\ref{eq4.1}). We remind that
\bea
\alpha=2\alpha_{+}a,\quad\beta=2\alpha_{+}b,\quad\gamma=2\alpha_{+}c\nn\\
\alpha'=2\alpha_{-}a,\quad\beta'=2\alpha_{-}b,\quad\gamma'=2\alpha_{-}c\label{eq4.5}
\eea
and, by the neutrality condition for $a,b,c$ in (\ref{eq4.1}),
\beq
l=\frac{m'+n'+p'-1}{2},\quad k=\frac{m+n+p-1}{2}\label{eq4.6}
\eeq
The expression in (\ref{eq4.4}) could additionally be symmetrized by transforming $\Gamma$'s with $-k$,
as follows:
\beq
\frac{\Gamma(x-k)}{\Gamma(1-x+k)}=\prod^{k}_{j=1}\frac{(-1)}{(x-j)^{2}}\times\frac{\Gamma(x)}{\Gamma(1-x)}\label{eq4.7}
\eeq
This gives:
\bea
<V_{c}(\infty)V_{b}(1)V_{a}(0)>\nn\\
=\rho^{-4lk}\times\prod^{l}_{i=1}\prod^{k}_{j=1}\frac{1}{(i\rho'-j)^{2}}\times\prod^{l}_{i=1}\frac{\Gamma(i\rho')}{\Gamma(1-i\rho')}\times\prod^{k}_{j=1}\frac{\Gamma(j\rho)}{\Gamma(1-j\rho)}\nn\\
\times\prod^{l-1}_{i=0}\prod^{k-1}_{j=0}\frac{1}{(\alpha'+i\rho'-j)^{2}(\beta'+i\rho'-j)^{2}(\gamma'+i\rho'-j)^{2}}\nn\\ 
\times\prod^{l-1}_{i=0}\frac{\Gamma(1+\alpha'+i\rho')\Gamma(1+\beta'+i\rho')\Gamma(1+\gamma'+i\rho')}{\Gamma(-\alpha'-i\rho')\Gamma(-\beta'-i\rho')\Gamma(-\gamma'-i\rho')}\nn\\
\times\prod^{k-1}_{j=0}\frac{\Gamma(1+\alpha+j\rho)\Gamma(1+\beta+j\rho)\Gamma(1+\gamma+j\rho)}{\Gamma(-\alpha-j\rho)\Gamma(-\beta-j\rho)\Gamma(-\gamma-j\rho)}\label{eq4.8}
\eea
Sometimes the formula (\ref{eq4.8}) is more convenient to make varions check.

We remind also that this same correlation function, expressed in terms of the function $\Upsilon_{M}(x)$, is of the form
\bea
<V_{c}(\infty)V_{b}(1)V_{a}(0)>\nn\\
=\frac{\Upsilon_{M}(a+b+c-2\alpha_{0})\Upsilon_{M}(-a+b+c)\Upsilon_{M}(a-b+c)\Upsilon_{M}(a+b-c)}{\Upsilon_{M}(2a)\Upsilon_{M}(2b)\Upsilon_{M}(2c)}\nn\\
\times(\rho)^{l(1-\rho')}\times(\rho')^{k(1-\rho)}\label{eq4.9}
\eea
-- comp.(\ref{eq3.39}), with $\rho$ factor expressed as in (\ref{eq3.42}). We remind that we are actually in the normalisation (\ref{eq4.3}) of the screening constants $\mu_{+}$, $\mu_{-}$.

Going back to the problem of normalisation of vertex operators $V_{a}$, $V_{b}$, $V_{c}$, we shall base our arguments on several simple examples.

It is shown in the Appendix B, by using the formulas in (\ref{eq4.8}) and (\ref{eq4.9}), that we have the following particular results:

1.
\beq
<III>=Z\label{eq4.10}
\eeq
where
\beq
Z=\frac{-\rho}{(1-\rho)^{2}}\gamma(\rho')\gamma(\rho)=\Upsilon(-2\alpha_{0})\rho^{\rho-\rho'}\label{eq4.11}
\eeq
$I$ is the identity operator, $I=1$. $\gamma(\rho)=\Gamma(\rho)/\Gamma(1-\rho)$. $Z$ could be considered as the partition function of the Coulomb gas, because the function $<III>=<1>$ is given by the Coulomb gas functional integral [7,8], without normalisation. So we note it as $Z$. The result (\ref{eq4.10}) is obtained in the Appendix B, both with (\ref{eq4.8}), by the analytic continuation of $<V_{c}V_{b}V_{a}>$ in its charges, $a\rightarrow 0$, $b\rightarrow 0$, $c\rightarrow 0$, and also with (\ref{eq4.9}).

2.
\beq
<I^{+}II>=1\label{eq4.12}
\eeq
Here $I^{+}(z,\bar{z})=V_{2\alpha_{0}}(z,\bar{z})$ is the conjugate identity operator.

3.
\beq
<I^{+}I^{+}I>=\frac{1}{Z}\label{eq4.13}
\eeq

4.
\beq
<I^{+}I^{+}I^{+}>=\frac{1}{Z^{2}}\label{eq4.14}
\eeq

5.
\beq
<I^{+}V_{1,2}V_{1,2}>=(N_{1,2})^{2}\label{eq4.15}
\eeq
where
\beq
(N_{1,2})^{2}=\frac{\gamma(2\rho-1)}{\gamma(\rho)}\label{eq4.16}
\eeq
is the "naive" norm squared of the operator $V_{1,2}$. More generally [11],
Section 9.1:
\bea
(N_{1,n})^{2}=\prod^{n-1}_{j=1}\frac{\Gamma(1-j\rho)\Gamma(-1+(1+j)\rho)}{\Gamma(j\rho)\Gamma(2-(1+j)\rho)}\nn\\
=\prod^{n-1}_{j=1}\frac{\gamma((1+j)\rho-1)}{\gamma(j\rho)}\label{eq4.17}
\eea

6.
\beq
<IV_{1,2}V_{1,2}>=Z (N_{1,2})^{2}\label{eq4.18}
\eeq

7.
\beq
<IV^{+}_{1,2}V_{1,2}>=1\label{eq4.19}
\eeq

8.
\beq
<I^{+}V^{+}_{1,2}V_{1,2}>=Z^{-1}\label{eq4.20}
\eeq

9.
\beq
<I^{+}V^{+}_{1,2}V^{+}_{1,2}>=Z^{-2}(N_{1,2})^{-2}\label{eq4.21}
\eeq

By comparing $<III>$ in (\ref{eq4.10}) and $<I^{+}II>$ in (\ref{eq4.12}) we have to  conclude that
\beq
I^{+}=\frac{1}{Z}I\label{eq4.22}
\eeq

Similarly, by comparing (\ref{eq4.18}) for $<IV_{1,2}V_{1,2}>$ and (\ref{eq4.19}) for $<IV^{+}_{1,2}V_{1,2}>$ 
we have to  conclude that
\beq
V^{+}_{1,2}=\frac{1}{Z(N_{1,2})^{2}}V_{1,2}\label{eq4.23}
\eeq

The identifications in (\ref{eq4.22}) and (\ref{eq4.23}) are not in the sense of the Coulomb gas theory, where these operators are different, but in the sense of the corresponding statistical model (matter) theory, where we assume that $V$ and $V^{+}$ represent the same statistical model operator, like spin in the $q$ - component Pots model, 
for general, real values of $q$. In the sense that the two operators, in the equalities (\ref{eq4.22}), (\ref{eq4.23}), should give the same correlation functions.

Next, if we assume, naturally, that the result in (\ref{eq4.10}) for $<III>$ 
is in fact the partition function of the Coulomb gas, then (\ref{eq4.10}) could be rewritten as:
\beq
<III>=Z\cdot\frac{<III>}{Z}=Z\cdot<<III>>\label{eq4.24}
\eeq
where
\beq
<<III>>=\frac{<III>}{Z} \label{eq4.25}
\eeq
is the properly normalised correlation function. 
Then the result in (\ref{eq4.10}) for $<III>$ implies that
\beq
<<III>>=1\label{eq4.26}
\eeq
which assumes that $I=1$ is the properly normalised identity operator,
\beq
N(I)=1\label{eq4.27}
\eeq
In this case, by (\ref{eq4.22}),
\beq
N(I^{+})=\frac{1}{Z} \label{eq4.28}
\eeq
This is consistent with (\ref{eq4.12}), (\ref{eq4.13}), (\ref{eq4.14}). For instance:
\bea
<I^{+}I^{+}I^{+}>=Z\cdot<<I^{+}I^{+}I^{+}>>\nn\\
=Z\cdot<<\frac{1}{Z^{3}}III>>=\frac{1}{Z^{2}}\label{eq4.29}
\eea

Next, the result in (\ref{eq4.15}) could be interpreted as:
\bea
<I^{+}V_{1,2}V_{1,2}>=\frac{1}{Z}<IV_{1,2}V_{1,2}>\nn\\ 
=<<IV_{1,2}V_{1,2}>>=<<V_{1,2}V_{1,2}>>\label{eq4.30}
\eea
since $I=1$. By (\ref{eq4.15})
\beq
<<V_{1,2}V_{1,2}>>=N^{2}_{1,2}\label{eq4.31}
\eeq
so that the "naive" norm of $V_{1,2}$ is in fact its actual norm:
\beq
N(V_{1,2})=N_{1,2}\label{eq4.32}
\eeq
$(N_{1,2})^{2}$ given by (\ref{eq4.16}).

Again, because of the relation (\ref{eq4.23}), the norm of the conjugate operator is given by:
\bea
N(V^{+}_{1,2})=\frac{1}{Z\cdot(N_{1,2})^{2}}\cdot N_{1,2}\nn\\
N(V^{+}_{1,2})=\frac{1}{Z\cdot N_{1,2}}\label{eq4.33}
\eea
The consistency with (\ref{eq4.18}) - (\ref{eq4.21}) could readily be verified.

Now, in general,
\beq
N(V_{a})=N_{a},\quad N(V^{+}_{a})=\frac{1}{Z\cdot N_{a}}\label{eq4.34}
\eeq
where $(N_{a})^{2}$ is given by:
\beq
(N_{a})^{2}=<I^{+}V_{a}V_{a}>\label{eq4.35}
\eeq

The "naive" norms of vertex operators $V_{a}$, $V^{+}_{a}$ have been defined in [11], Section 9.1. They differ from the actual norms in (\ref{eq4.34}) 
by the absence of the partition function $Z$, of the Coulomb gas, 
in the norm of $V^{+}_{a}$, $N(V^{+}_{a})_{naive}=1/N_{a}$. 

By the formula (\ref{eq4.9}) we obtain:
\bea
<I^{+}V_{a}V_{a}>\nn\\
=\frac{\Upsilon_{M}(2a)\Upsilon_{M}(2\alpha_{0})\Upsilon_{M}(2\alpha_{0})\Upsilon_{M}(2a-2\alpha_{0})}{\Upsilon_{M}(2a)\Upsilon_{M}(2a)\Upsilon_{M}(4\alpha_{0})}\nn\\
\times\rho^{(n'-1)(1-\rho')}\times(\rho')^{(n-1)(1-\rho)}\label{eq4.36}
\eea
Here we assume that $a=\alpha_{n',n}$, and then, by (\ref{eq4.6}) $l=n'-1$, $k=n-1$. Since $\Upsilon_{M}(2\alpha_{0})=1$, $\Upsilon_{M}(x)=\Upsilon_{M}(2\alpha_{0}-x)$, Appendix B, we obtain:
\beq
(N_{a})^{2}=\frac{\Upsilon_{M}(2a-2\alpha_{0})}{\Upsilon_{M}(2a)\Upsilon_{M}(-2\alpha_{0})}\times\rho^{(n'-1)(1-\rho')}\times(\rho')^{(n-1)(1-\rho)}\label{eq4.37}
\eeq

Consistency could readily be checked (Appendix B) that:
\beq
(N(V^{+}_{a}))^{2}=(N_{a^{+}})^{2}
=<I^{+}V_{a^{+}}V_{a^{+}}>=<I^{+}V_{2\alpha_{0}-a}V_{2\alpha_{0}-a}>
=\frac{1}{Z^{2}(N_{a})^{2}}\label{eq4.38}
\eeq
-- consistent with (\ref{eq4.34}). We could conclude that the formula (\ref{eq4.37}), for the norm squared of the Coulomb gas vertex operator $V_{a}$, is perfectly general. Though still limited to the degenerate values of $a$, $a=\alpha_{n',n}$, 
as we  are still working with the formula (\ref{eq4.9}) 
which contains the $\rho$ factors, in which the numbers $l,k$ are still present.

We shall now define the correlation function, properly normalised by $\frac{1}{Z}$,  
and which is defined for the  normalised operators (\ref{eq3.47}). We find:
\bea
<<\Phi_{c}(\infty)\Phi_{b}(1)\Phi_{a}(0)>>
=\frac{1}{Z}<V_{c}(\infty)V_{b}(1)V_{a}(0)>\cdot \frac{1}{N_{a}N_{b}N_{c}}\nn\\
=\frac{ \rho^{-\rho+\rho'}}{\Upsilon_{M}(-2\alpha_{0})}
\frac{\Upsilon_{M}(a+b+c-2\alpha_{0})\Upsilon_{M}(-a+b+c)\Upsilon_{M}(a-b+c)\Upsilon_{M}(a+b-c)}{\Upsilon_{M}(2a)\Upsilon_{M}(2b)\Upsilon_{M}(2c)}\nn\\
\rho^{l(1-\rho')}(\rho')^{k(1-\rho)} [\frac{\Upsilon_{M}(2a)\Upsilon(-2\alpha_{0})}{\Upsilon_{M}(2a-2\alpha)}\times\frac{\Upsilon_{M}(2b)\Upsilon_{M}(-2\alpha_{0})}{\Upsilon_{M}(2b-2\alpha_{0})}\times\frac{\Upsilon_{M}(2c)\Upsilon_{M}(-2\alpha_{0})}{\Upsilon_{M}(2c-2\alpha_{0})}]^{1/2}\nn\\
(\rho)^{-\frac{m'+n'+p'-3}{2}(1-\rho')}(\rho')^{-\frac{m+n+p-3}{2}(1-\rho)}\label{eq4.39}
\eea
We have assumed that $a=\alpha_{m',m}$, $b=\alpha_{n',n}$, $c=\alpha_{p',p}$.
We obtain:
\bea
\frac{\Upsilon_{M}(a+b+c-2\alpha_{0})\Upsilon_{M}(-a+b+c)\Upsilon_{M}(a-b+c)\Upsilon_{M}(a+b-c)}{\sqrt{\Upsilon_{M}(2a)\Upsilon_{M}(2a-2\alpha_{0})\times\Upsilon_{M}(2b)\Upsilon_{M}(2b-2\alpha_{0})\times\Upsilon_{M}(2c)\Upsilon_{M}(2c-2\alpha_{0})}}\nn\\
\times\sqrt{\Upsilon_{M}(-2\alpha_{0})}\times\rho^{-\rho+\rho'}\times\rho^{l(1-\rho')}(\rho')^{k(1-\rho)}\nn\\
\times(\rho)^{-l(1-\rho')+(1-\rho')}\times(\rho')^{-k\cdot(1-\rho')+(1-\rho)}\label{eq4.40}
\eea
We have used the formulas (\ref{eq4.6}) for $l$ and $k$. We observe that all the $\rho$-factors get cancelled and we find, finally, the normalised 3-point function 
for the normalised operators in the form:
\bea
<<\Phi_{c}(\infty)\Phi_{b}(1)\Phi_{a}(0)>>\nn\\
=\frac{\Upsilon_{M}(a+b+c-2\alpha_{0})\Upsilon_{M}(-a+b+c)\Upsilon_{M}(a-b+c)\Upsilon_{M}(a+b-c)\sqrt{\Upsilon_{M}(-2\alpha_{0})}}{\sqrt{\Upsilon_{M}(2a)\Upsilon_{M}(2a-2\alpha_{0})\times\Upsilon_{M}(2b)\Upsilon_{M}(2b-2\alpha_{0})\times\Upsilon_{M}(2c)\Upsilon_{M}(2c-2\alpha_{0})}}\label{eq4.41}
\eea
In this formula everything is expressed, analytically, in terms of charges $a,b,c$, so that we can continue the formula to the general, continuons values of charges.

The expression (\ref{eq4.41}) is the formula (5.1) of [6], obtained there differently.
Our point is that we have derived everything, by a series of analytic continuations, 
from the general 3-point function of the minimal model [9].

\vskip1.5cm

\section{Analytic continuation of the function $<<\Phi_{c}\Phi_{b}\Phi_{a}>>$
towards the 3-point function of Liouville.}

The 3-point function $<<\Phi_{c}\Phi_{b}\Phi_{a}>>$, which have been defined 
in the Section 4, could further be analytically continued to give the 3-point function 
of the Liouville model. We have to continue the charges
\beq 
a\rightarrow -ia,\quad b\rightarrow -ib, \quad c\rightarrow -ic\label{eq5.1}
\eeq
so that the vertex operators $V_{a}=e^{ia\varphi}$, $V_{b}=e^{ib\varphi}$, $V_{c}=e^{ic\varphi}$, eventually normalised, $V_{a}\rightarrow\Phi_{a}=\frac{1}{N_{a}}V_{a}$, etc., would go to the Liouville model vertex operators $e^{a\varphi}$, $e^{b\varphi}$, $e^{c\varphi}$. The central charge parameter $h(=\alpha_{+})$ of the corresponding conformal theory, 
has also to be continued:
\beq
h\rightarrow -ih\label{eq5.2}
\eeq

Since the function $<<\Phi_{c}\Phi_{b}\Phi_{a}>>$ is expressed as a product of $\Upsilon_{M}(x,h)$ functions, eq.(\ref{eq4.41}), we have to continue first the function $\Upsilon_{M}(x,h)$, towards $\Upsilon_{M}(-ix,-ih)$, and then we shall have to construct with it the analytic continuation of the 3-point function $<<\Phi_{c}\Phi_{b}\Phi_{a}>>$.

\numberwithin{equation}{subsection}

\vskip1cm

\subsection{Analytic continuation of the function $\Upsilon_{M}(x,h).$}

For convenience, we reproduce here the integral definition of $\log \Upsilon_{M}(x,h)$, eq.(\ref{eq3.28}):
\bea
\log\Upsilon_{M}(x,h)=\int_{0}^{\infty}\frac{dt}{t}\{(\alpha_{0}-x)^{2}e^{-t}-\frac{\sinh^{2}[(\alpha_{0}-x)\frac{t}{2}]}{\sinh\frac{th}{2}\cdot\sinh\frac{t}{2h}}\},\nn\\
\alpha_{0}=\frac{h}{2}-\frac{1}{2h}.\label{eq5.3}
\eea
We shall put
\beq
x=\tilde{x}e^{-i\theta},\quad h=\tilde{h}e^{-i\theta}; \quad \theta: 0\rightarrow\frac{\pi}{2}\label{eq5.4}
\eeq
where $\tilde{x}$, $\tilde{h}$ are considered to be real, positives, for the moment. (\ref{eq5.3}) takes the form:
\bea
\log\Upsilon_{M}(\tilde{x}e^{-i\theta},\tilde{h}e^{-i\theta})\nn\\
=\int_{0}^{\infty}\frac{dt}{t}\{(\frac{\tilde{h}}{2}e^{-i\theta}-\frac{1}{2\tilde{h}}e^{i\theta}-\tilde{x}e^{-i\theta})^{2}e^{-t}\nn\\
-\frac{\sinh[(\frac{\tilde{h}}{2}e^{-i\theta}-\frac{1}{2\tilde{h}}e^{i\theta}-\tilde{x}e^{-i\theta})\frac{t}{2}]}{\sinh(\frac{t\tilde{h}}{2}e^{-i\theta})\sinh(\frac{t}{2\tilde{h}}e^{i\theta})}\}\label{eq5.5}
\eea
In Fig.1 are shown the poles of the expression under the integral in (\ref{eq5.5}) considered as a function in the complex plane of $t$. We observe that there 
are no poles at $t=0$.

\begin{figure}
\begin{center}
\epsfxsize=300pt\epsfysize=210pt{\epsffile{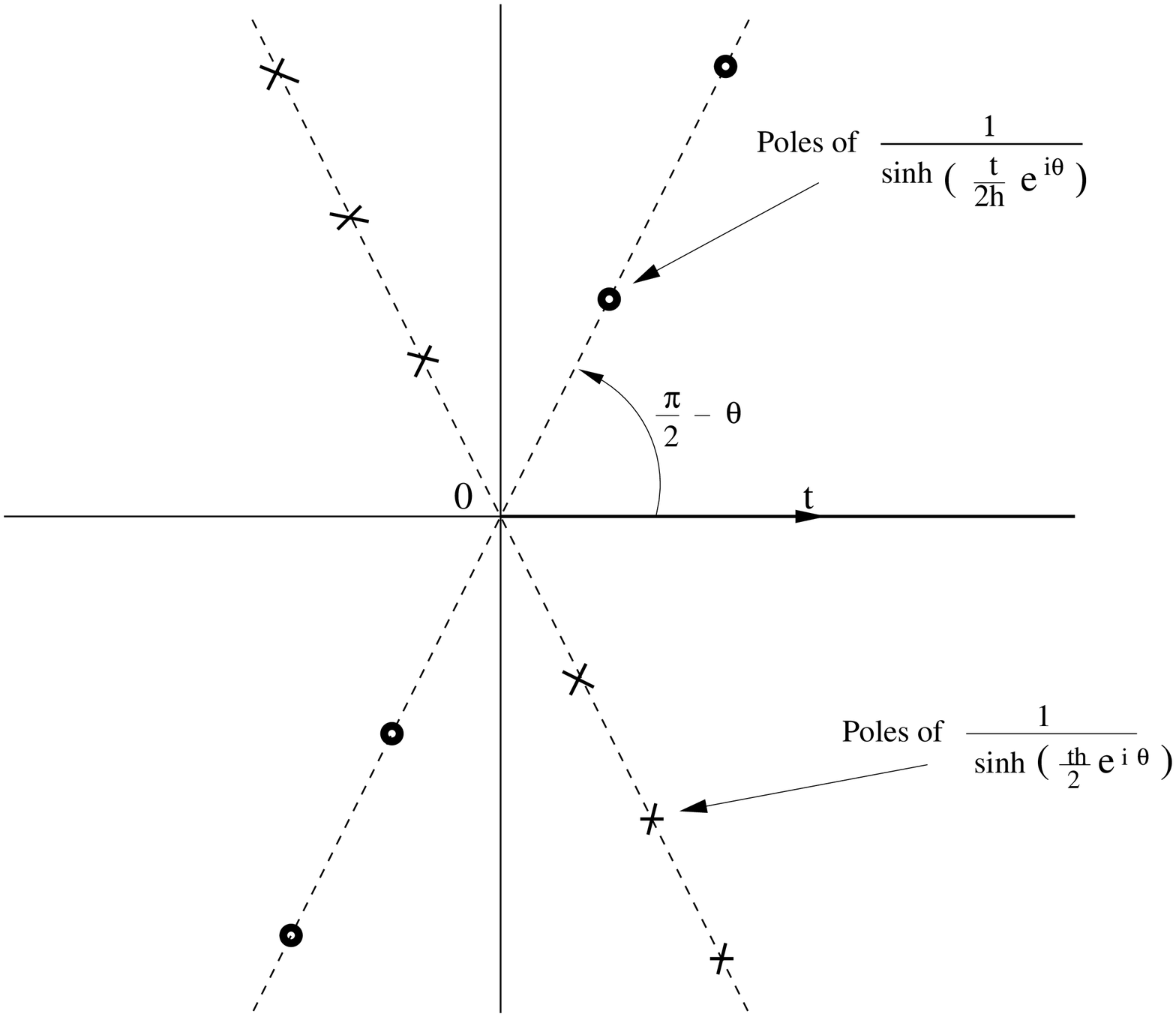}}
\caption{The start of the analytic continuation of the function 
$\Upsilon_{M}$:  $\Upsilon_{M}(x,h) \rightarrow
\Upsilon_{M}(\tilde{x} e^{-i\theta},\tilde{h} e^{-i\theta})$. 
The plane of the figure is the complex plane of $t$.
\label{fffig1}
}
\end{center}
\end{figure}

The poles are due to the factors
\beq
\frac{1}{\sinh(\frac{t\tilde{h}}{2}e^{-i\theta})},\quad\frac{1}{\sinh(\frac{t}{2\tilde{h}}e^{i\theta})}\label{eq5.6}
\eeq
As $\theta\rightarrow\frac{\pi}{2}$, the poles approach the real axes, the integration line of $t$. For $\theta=\frac{\pi}{2}$ the poles put themselves on the real axes and the integration line, over $t$, gets deformed accordingly,  to avoid poles, Fig.2.

\begin{figure}
\begin{center}
\epsfxsize=300pt\epsfysize=190pt{\epsffile{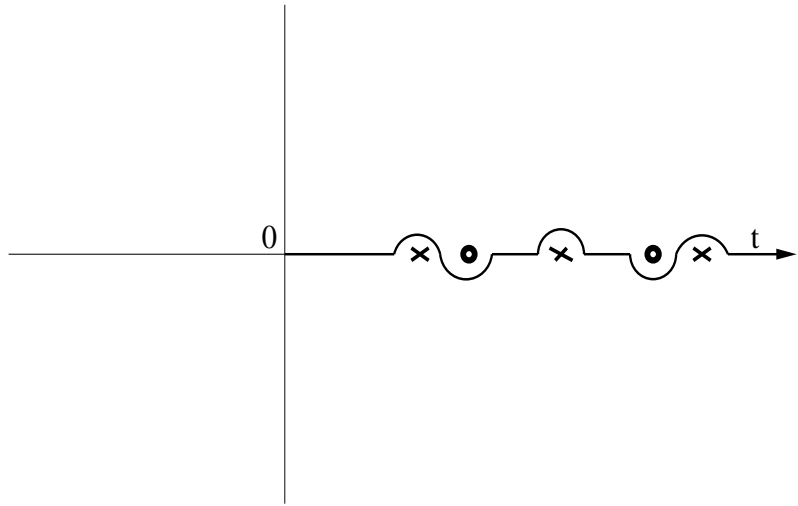}}
\caption{The Fig.1 when $\theta$ becomes equal to  $\pi/2$, 
$\Upsilon_{M}(\tilde{x} e^{-i\theta},\tilde{h} e^{-i\theta})
\rightarrow \Upsilon_{M}(-i\tilde{x},-i\tilde{h})$.
\label{ffig2}
}
\end{center}
\end{figure}

Next, the contour of integration could be deformed as shown in Fig.3. 

\begin{figure}
\begin{center}
\epsfxsize=300pt\epsfysize=210pt{\epsffile{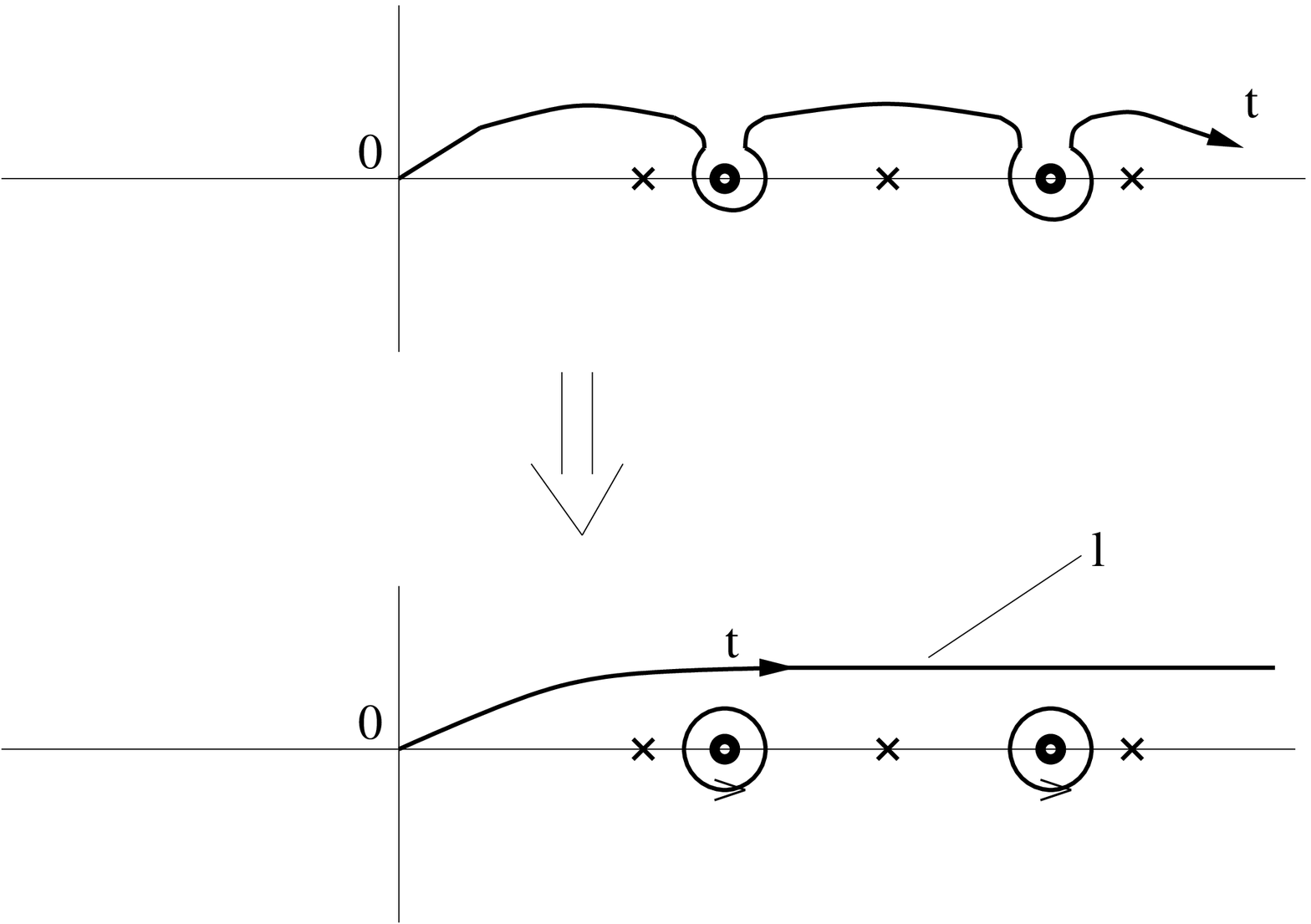}}
\caption{Successive deformations of the integration contour over t, 
in the integral  for $\log \Upsilon_{M}(-i\tilde{x}, -i\tilde{h})$.
\label{ffig3}
}
\end{center}
\end{figure}

Accordingly, the integration breaks into two part: the sum of integrals around poles, the first part, and the integral along the line $l$, Fig.3, the second part. We shall calculate them successively.
\beq
Poles=\sum_{n=1}^{\infty}\oint_{C_{n}}\frac{dt}{t}(-\frac{\sinh^{2}[(\frac{-i\tilde{h}}{2}-\frac{i}{2\tilde{h}}+i\tilde{x})\frac{t}{2}]}{\sinh(-\frac{it\tilde{h}}{2})\cdot\sinh(i\frac{t}{2\tilde{h}})})\label{eq5.7}
\eeq
$C_{n}$ is a small closed contour around $t_{n}=2\pi\tilde{h}\cdot n$, the pole due to the factor $1/\sinh(\frac{it}{2\tilde{h}})$. We get:
\bea
Poles=\sum_{n=1}^{\infty}\oint_{C_{n}}\frac{dt}{t}(-\frac{e^{(-\frac{i\tilde{h}}{2}-\frac{i}{2\tilde{h}}
+i\tilde{x})2\pi\tilde{h}n}+c.c.-2}{2(e^{\frac{-i\tilde{h}}{2}2\pi\tilde{h}n}-e^{\frac{i\tilde{h}}{2}2\pi\tilde{h}n})i(-1)^{n}\cdot\frac{1}{2\tilde{h}}(t-2\pi\tilde{h}n)})\nn\\
=\sum_{n=1}^{\infty}\frac{2\pi i}{2\pi\tilde{h}n}\cdot\frac{i}{2}(-1)^{n}\cdot2\tilde{h}
\times\frac{e^{-i\pi\tilde{h}^{2}n-i\pi n+2\pi i\tilde{h}\tilde{x}n}+c.c.-2}{(e^{-i\pi\tilde{h}^{2}n}-e^{i\pi\tilde{h}^{2}n})}\nn\\
=-\sum^{\infty}_{n=1}\frac{1}{n}\cdot\frac{e^{-i\pi\tilde{h}^{2}n+2\pi i\tilde{h}\tilde{x}n}+e^{i\pi\tilde{h}^{2}n-2\pi i\tilde{h}\tilde{x}n}-2\cdot(-1)^{n}}{(e^{-i\pi\tilde{h}^{2}n}-e^{i\pi\tilde{h}^{2}n})}\label{eq5.8}
\eea
\beq
Poles=-\sum_{n=1}^{\infty}\frac{1}{n}\cdot\frac{e^{2\pi i\tilde{h}\tilde{x}n}+q^{2n}\cdot e^{-2\pi i\tilde{h}\tilde{x}n}-2(-1)^{n}\cdot q^{n}}{1-q^{2n}}\label{eq5.9}
\eeq
with
\beq
q=e^{i\pi \tilde{h}^{2}}\label{eq5.10}
\eeq
The series in(\ref{eq5.9}) is almost that for the $\log$ of ratio of two $\vartheta$-functions:
\beq
-\sum^{\infty}_{n=1}\frac{1}{n}\cdot\frac{e^{2iun}+q^{2n}e^{-2iun}-2(-1)^{n}q^{n}}{1-q^{n}}
=-\frac{1}{4}\log q-i(\frac{\pi}{2}-u)+\log\frac{\vartheta_{1}(u,q)}{\vartheta_{3}(0,q)}\label{eq5.11}
\eeq
This formula is obtained in the Appendix C. Using (\ref{eq5.11}), the result (\ref{eq5.9}), for $Poles$, could be given as:
\beq
Poles=-\frac{1}{4}\log q-i(\frac{\pi}{2}-\pi\tilde{h}\tilde{x})+\log\frac{\vartheta_{1}(\pi\tilde{h}\tilde{x},q)}{\vartheta_{3}(0,q)}, \quad \,\,\,
q=e^{i\pi\tilde{h}^{2}}\label{eq5.12}
\eeq

The second part of (\ref{eq5.5}) (with $\theta=\frac{\pi}{2}$) is given by the integral along the line $l$, Fig.3:
\bea
I_{l}=\int^{\infty}_{0}\frac{dt}{t}\{(-\frac{i\tilde{h}}{2}-\frac{i}{2\tilde{h}}+i\tilde{x})^{2}e^{-t}\nn\\
-\frac{\sinh^{2}[(-\frac{i\tilde{h}}{2}-\frac{i}{2\tilde{h}}+i\tilde{x})\frac{t}{2}]}{\sinh(\frac{-i\tilde{h}}{2}t)\cdot\sinh(\frac{i}{2\tilde{h}}t)}\}\label{eq5.13})
\eea
We shall break it into pieces as:
\bea
I_{l}=\lim_{\epsilon\rightarrow 0}\{-\int_{\epsilon}^{\infty}\frac{dt}{t}(\frac{\tilde{h}}{2}+\frac{1}{2\tilde{h}}-\tilde{x})^{2}e^{-t}\nn\\
+\int_{C_{\epsilon}}\frac{dt}{t}\frac{(\frac{\tilde{h}}{2}+\frac{1}{2\tilde{h}}-\tilde{x})^{2}\cdot\frac{t^{2}}{4}}{\frac{\tilde{h}}{2}t\cdot\frac{1}{2\tilde{h}}t}\nn\\
+\int_{\epsilon}^{\infty}\frac{d\tilde{t}}{\tilde{t}}\frac{\sinh^{2}[(\frac{\tilde{h}}{2}+\frac{1}{2\tilde{h}}-\tilde{x})\frac{\tilde{t}}{2}]}{\sinh(\frac{\tilde{h}\tilde{t}}{2})\cdot\sinh(\frac{\tilde{t}}{2\tilde{h}})}\}\label{eq5.14}
\eea
In the above, we have introduced the lower limit of integration $\epsilon$, with the limit $\epsilon\rightarrow 0$, because the integrals in (\ref{eq5.14}), taking separately, are divergent at $t\rightarrow 0$.
Then we kept the integration contour as it is, $t:\epsilon\rightarrow\infty$, in the first integral. For this part, there are no poles and the integration line $l$, Fig.3, could be put back on the real axes. While for the integral of the second term in (\ref{eq5.13}), the part which contains
poles in the vicinity of $l$, Fig.3, we have turned the contour of integration, the line $l$, towards the imaginary axes, as is shown in Fig.4. 

\begin{figure}
\begin{center}
\epsfxsize=300pt\epsfysize=230pt{\epsffile{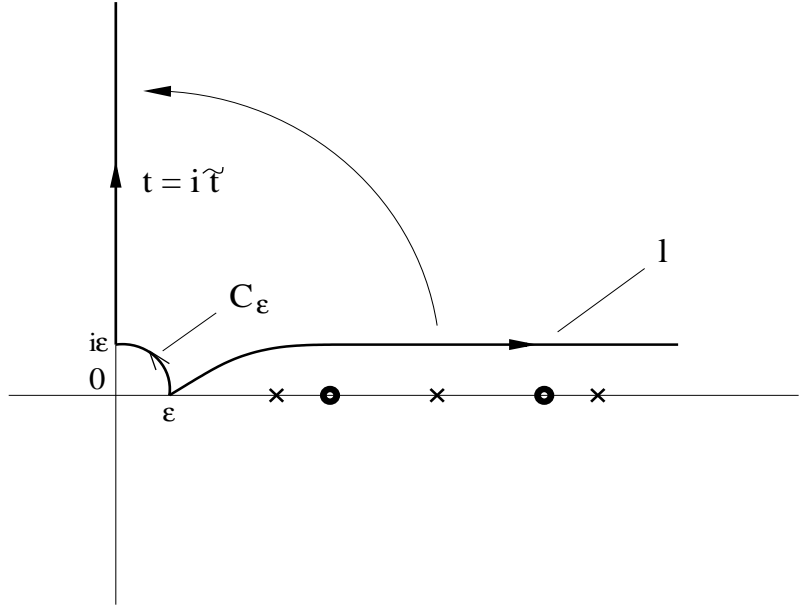}}
\caption{Final deformation of the integration contour along the line $l$,
in the integral $I_{l}$.
\label{ffig4}
}
\end{center}
\end{figure}

This contour then break in two, the small contour $C_{\epsilon}$ around the origin, and the integral along the imaginary axes, for which we have changed the variable $t=i\tilde{t}$, Fig.4. This last integral is the third one in (\ref{eq5.14}). We shall drop the tildes, of $\tilde{t}$, in third term of (\ref{eq5.14}), in the following. In the second integral, over $C_{\epsilon}$ in (\ref{eq5.14}), we have already developed the $\sinh$'s in the numerator and in the denominator. This integral is equal to
\beq
i\frac{\pi}{2}(\beta_{0}-\tilde{x})^{2}\label{eq5.15}
\eeq
where
\beq
\beta_{0}=\frac{\tilde{h}}{2}+\frac{1}{2\tilde{h}}\label{eq5.16}
\eeq
The first and the third integrals could be put together and the limit $\epsilon\rightarrow 0$ could be lifted. Altogether, we get
\beq
I_{l}=i\frac{\pi}{2}(\beta_{0}-\tilde{x})^{2}-\log\Upsilon_{L}(\tilde{x},\tilde{h})\label{eq5.17}
\eeq
Here $\Upsilon_{L}(\tilde{x},\tilde{h})$ is the $\Upsilon$ function for the Liouville model, which was introduced in [5]:
\beq
\log\Upsilon_{L}(\tilde{x},\tilde{h})=\int_{0}^{\infty}\frac{dt}{t}\{(\beta_{0}-\tilde{x})^{2}e^{-t}-\frac{\sinh^{2}[(\beta_{0}-\tilde{x})\frac{\tilde{t}}{2}]}{\sinh(\frac{\tilde{h}t}{2})\cdot\sinh(\frac{t}{2\tilde{h}})}\}\label{eq5.18}
\eeq
$\beta_{0}$ is given by(\ref{eq5.16}).

Now, putting together the two parts of (\ref{eq5.5}) (with $\theta=\frac{\pi}{2}$), 
the $Poles$, eq.(\ref{eq5.12}), and $I_{l}$, eq.(\ref{eq5.17}), we obtain:
\bea
\log\Upsilon_{M}(-i\tilde{x},-i\tilde{h})\nn\\
=-\frac{1}{4}\log q-i(\frac{\pi}{2}-\pi\tilde{h}\tilde{x})+\log\frac{\vartheta_{1}(\pi\tilde{h}\tilde{x},q)}{\vartheta_{3}(0,q)}\nn\\
+i\frac{\pi}{2}(\beta_{0}-\tilde{x})^{2}-\log\Upsilon_{L}(\tilde{x},\tilde{h})\label{eq5.19}
\eea
or
\beq
\Upsilon_{M}(-i\tilde{x},-i\tilde{h})\Upsilon_{L}(\tilde{x},\tilde{h})=\frac{1}{q^{1/4}}e^{i\frac{\pi}{2}(\alpha_{0}+\tilde{x})^{2}}\times\frac{\vartheta_{1}(\pi\tilde{h}\tilde{x},q)}{\vartheta_{3}(0,q)}\label{eq5.20}
\eeq
We remind that
\beq
\alpha_{0}=\frac{\tilde{h}}{2}-\frac{1}{2\tilde{h}},\quad\beta_{0}=\frac{\tilde{h}}{2}+\frac{1}{2\tilde{h}},\quad q=e^{i\pi\tilde{h}^{2}}\label{eq5.21}
\eeq
When passing from (\ref{eq5.19}) to (\ref{eq5.20}) we have grouped together two terms in (\ref{eq5.19}):
\bea
i\frac{\pi}{2}(\beta_{0}-\tilde{x})^{2}-i(\frac{\pi}{2}-\pi\tilde{h}x)
=i\frac{\pi}{2}(\beta_{0}^{2}-2\beta_{0}\tilde{x}+\tilde{x}^{2}-1+2\tilde{h}x)\nn\\
=i\frac{\pi}{2}(\frac{\tilde{h}^{2}}{4}+\frac{1}{2}+\frac{1}{4\tilde{h}^{2}}-(\tilde{h}+\frac{1}{\tilde{h}})\tilde{x}+\tilde{x}^{2}-1+2\tilde{h}\tilde{x})\nn\\
=i\frac{\pi}{2}(\frac{\tilde{h}^{2}}{4}-\frac{1}{2}+\frac{1}{4\tilde{h}^{2}}+(\tilde{h}-\frac{1}{\tilde{h}})\tilde{x}+\tilde{x}^{2})
=i\frac{\pi}{2}(\alpha_{0}^{2}+2\alpha_{0}\tilde{x}+\tilde{x}^{2})=i\frac{\pi}{2}(\alpha_{0}+\tilde{x})^{2}\label{eq5.21a}
\eea

The relation (\ref{eq5.20}) is the formula (6.2) of [6], derived there by different methods.

\vskip0.5cm

One comment is in order, with respect to our derivation of the formula (\ref{eq5.20}).

The $\vartheta$ functions in (\ref{eq5.20}) are not defined for $q=e^{i\pi\tilde{h}^{2}}$ 
with $\tilde{h}$ real. We need to  have
\beq
Im\,\tilde{h}^{2}>0 \label{eq5.22}
\eeq
and then $\tilde{h}$ ought be in the sector $S$, Fig.5. 
\begin{figure}
\begin{center}
\epsfxsize=300pt\epsfysize=130pt{\epsffile{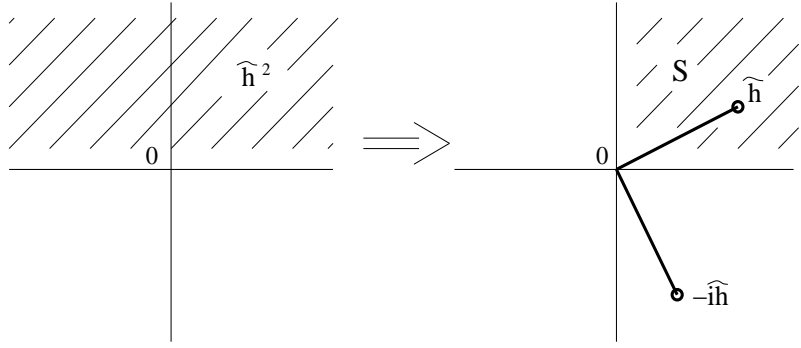}}
\caption{Domains in the complex planes of $\tilde{h}^{2}$ and of $\tilde{h}$
where the functions $\vartheta(x,q)$, with $q=e^{i\tilde{h}^{2}}$,
are well defined.
\label{ffig5}
}
\end{center}
\end{figure}

Also the argument $-i\tilde{h}$ of
$\Upsilon_{M}(-i\tilde{x},-i\tilde{h})$ in (\ref{eq5.20}) will, in this case, 
be in the sector below, Fig.5, such that the poles of the factors
\beq
\frac{1}{\sinh(\frac{-i\tilde{h}t}{2})\sinh(\frac{it}{2\tilde{h}})}\label{eq5.23}
\eeq
in the integral form of $\log\Upsilon_{M}(-i\tilde{x},-i\tilde{h})$, would stay away, still, from the integration line over $t$ (real axes): $-i\tilde{h}$ should stay away, to the right, from the lower part of the imaginary axes, for the integral representation of $\log\Upsilon_{M}(-i\tilde{x},-i\tilde{h})$ were well defined.

In summary, $\tilde{h}$ should be in the sector $S$, Fig.5, of its complex plane, 
for our analytic continuation had to make sense.

This implies that, at the start of our analytic continuation, 
$h=\tilde{h}$ (eq.(\ref{eq5.4}) for $\theta=0$) had to have a "small" imaginary part, positive. This implies in turn that our figures should slightly be deformed, as is indicated in Fig.6, Fig.7. Otherwise, the derivation stays as has been presented above, though in a somewhat (artificially) simplified context.

By the way, the fact that $\log\Upsilon_{M}(-i\tilde{x}, -i\tilde{h})$, in (\ref{eq5.20}), is well defined, by its integral, as was discussed above, is seen by the first figure in Fig.6: the poles stay away from the initial integration line, the real axes of $t$.

\begin{figure}
\begin{center}
\epsfxsize=300pt\epsfysize=300pt{\epsffile{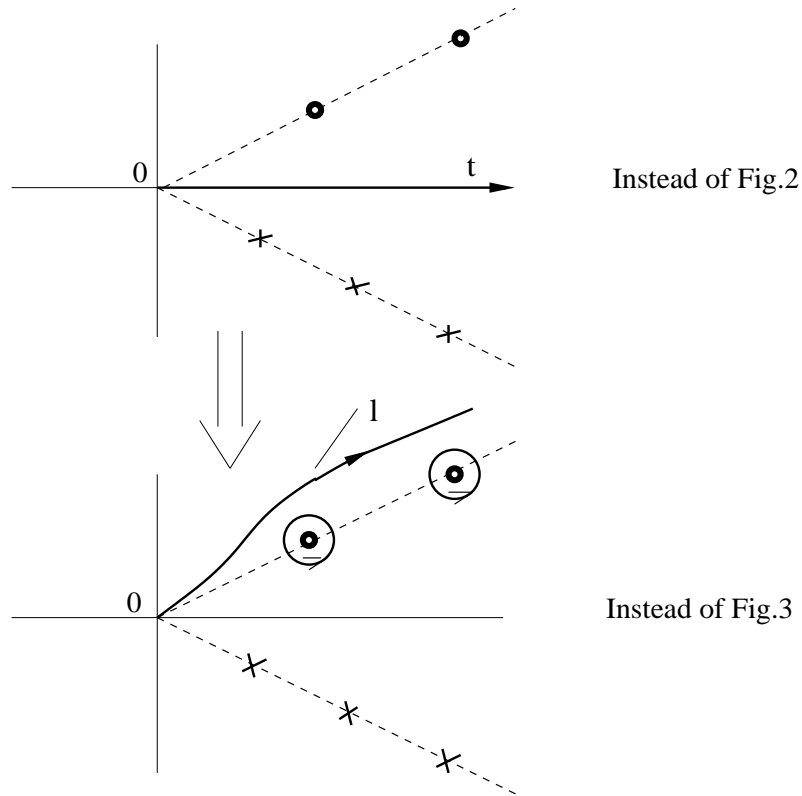}}
\caption{Modifications of Figures 2 and 3 when $\tilde{h}$ is complex, i.e.
$h=\tilde{h}e^{-i\theta}$ is complex initially, at the start 
of the analytic continuation, and $0<arg \,\tilde{h}<\pi/2$.
\label{ffig6}
}
\end{center}
\end{figure}
\begin{figure}
\begin{center}
\epsfxsize=300pt\epsfysize=250pt{\epsffile{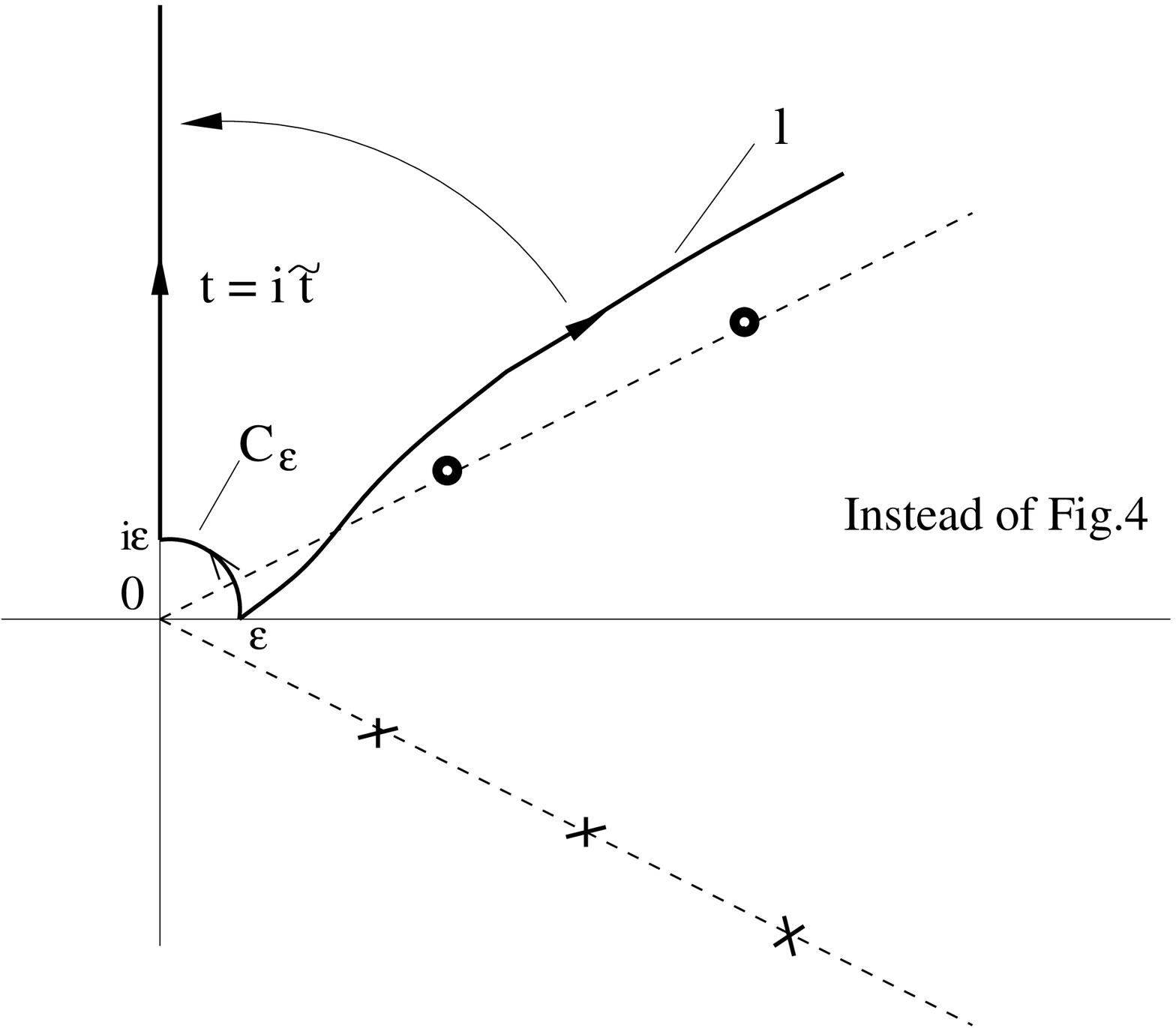}}
\caption{The final deformation of the contour of integration along the line $l$,
in the case when $\tilde{h}$ is complex.
\label{ffig7}
}
\end{center}
\end{figure}

\vskip1cm

\subsection{Analytic continuation of the function $<<\Phi_{c}\Phi_{b}\Phi_{a}>>$.}

The properly normalised 3-point function $<<\Phi_{c}\Phi_{b}\Phi_{a}>>$ 
in (\ref{eq4.41}) is all expressed in terms of the function $\Upsilon_{M}(x,h)$. 
To continue it to the Liouville sector we just have to replace $\Upsilon_{M}(x,h)$ by its analytically continued form, given by the formula (\ref{eq5.20}):
\beq
\Upsilon_{M}(-ix,-ih)=\frac{1}{q^{1/4}}\cdot\frac{1}{\Upsilon_{L}(x,h)}e^{i\frac{\pi}{2}(\alpha_{0}+x)^{2}}\times\frac{\vartheta_{1}(\pi hx,q)}{\vartheta_{3}(0,q)}\label{eq5.24}
\eeq
$q=e^{i\pi h^{2}}$. We have suppressed the "tildes" of $\tilde{x}$ and $\tilde{h}$, which served us in the subsection 5.1 for the presentation purposes of the analytic continuation.

As was discussed above, at the end of the subsection 5.1, for the formula (\ref{eq5.24}) to be valid, $h$ have to be complex, with
\beq
0<\arg h<\frac{\pi}{2}\label{eq5.25}
\eeq
-- $h$ have to be in the Sector S, Fig.5, $\tilde{h}\equiv h$ in (\ref{eq5.24}) and in Fig.5. 

Replacing every $\Upsilon_{M}$ in (\ref{eq4.41}) by its analytically continued form, eq.(\ref{eq5.24}), \\ we obtain:
\bea
<<\Phi_{c}(\infty)\Phi_{b}(1)\Phi_{a}(0)>>_{continued}\nn\\
=\sqrt{\Upsilon_{M}(-i(-2\beta_{0}),-ih)}\times [\Upsilon_{M}(-i(a+b+c-2\beta_{0}),-ih)\nn\\ 
\times\Upsilon_{M}(-i(-a+b+c),-ih)\Upsilon_{M}(-i(a-b+c),-ih)\Upsilon_{M}(-i(a+b-c),-ih)]\nn\\
/[\Upsilon_{M}(-2ia,-ih)\Upsilon_{M}(-i(2a-2\beta_{0}),-ih)
\Upsilon_{M}(-i2b,-ih)\nn\\ \times \Upsilon_{M}(-i(2b-2\beta_{0}),-ih)
\Upsilon_{M}(-i2c,-ih)\Upsilon_{M}(-i(2c-2\beta_{0}),-ih)]^{1/2}\nn\\
=\frac{1}{q^{1/4}}\times\frac{1}{\vartheta_{3}(0,q)}\times\sqrt{\Upsilon_{M}(-i(-2\beta_{0}),-ih)}\nn\\
\times\exp\{i\frac{\pi}{2}[(\alpha_{0}+a+b+c-2\beta_{0})^{2}+(\alpha_{0}-a+b+c)^{2}+(\alpha_{0}+a-b+c)^{2}\nn\\
+(\alpha_{0}+a+b-c)^{2}
-\frac{1}{2}(\alpha_{0}+2a)^{2}-\frac{1}{2}(\alpha_{0}+2b)^{2}
-\frac{1}{2}(\alpha_{0}+2c)^{2}\nn\\
-\frac{1}{2}(\alpha_{0}+2a-2\beta_{0})^{2}-\frac{1}{2}(\alpha_{0}+2b-2\beta_{0})^{2}-\frac{1}{2}(\alpha_{0}+2c-2\beta_{0})^{2}]\}\times\nn\\
\frac{\sqrt{\Upsilon_{L}(2a,h)\Upsilon_{L}(2a-2\beta_{0},h)\Upsilon_{L}(2b,h)
\Upsilon_{L}(2b-2\beta_{0},h)\Upsilon_{L}(2c,h)\Upsilon_{L}(2c-2\beta_{0},h)}}
{\sqrt{\vartheta_{1}(\pi h2a)\vartheta_{1}(\pi h(2a-2\beta_{0}))\vartheta_{1}(\pi h2b)
\vartheta_{1}(\pi h(2b-2\beta_{0}))\vartheta_{1}(\pi h2c)
\vartheta_{1}(\pi h(2c-2\beta_{0}))}}\nn\\
\times\frac{\vartheta_{1}(\pi h(a+b+c-2\beta_{0}))\vartheta_{1}(\pi h(-a+b+c))
\vartheta_{1}(\pi h(a-b+c))\vartheta_{1}(\pi h(a+b-c))}
{\Upsilon_{L}(a+b+c-2\beta_{0},h)\Upsilon_{L}(-a+b+c,h)\Upsilon_{L}(a-b+c,h)
\Upsilon_{L}(a+b-c,h)}\label{eq5.26}
\eea
We have suppressed the argument $q$ in  the functions $\vartheta_{1}$ 
($\vartheta_{1}(\pi h2a,q) \rightarrow \vartheta_{1}(\pi h2a)$, etc.) 
to compactify the expression a little bit.

With some simple algebra one can reduce $\exp\{i\frac{\pi}{2}[...]\}$ in the r.h.s. of (\ref{eq5.26}) to
\beq
\exp\{i\frac{\pi}{2}[-2\beta^{2}_{0}+2\alpha_{0}\beta_{0}+\alpha_{0}^{2}]\}\label{eq5.27}
\eeq
and one can check, with some manipulations for  $\Upsilon_{M}(-i(-2\beta_{0}),-ih)$, 
that \\ $\sqrt{\Upsilon_{M}(-i(-2\beta_{0}),-ih)}$ is equal to 
\beq
\frac{1}{4\alpha_{0}\beta_{0}}\sqrt{\frac{-1}{\Upsilon_{M}(-2\alpha_{0},h)}}e^{-i2\pi\alpha_{0}\beta_{0}}\label{eq5.28}
\eeq
Finally, the whole expression, in the r.h.s. of (\ref{eq5.26}), preceding the part with $\Upsilon$ functions, could be reduced to:
\beq
\frac{e^{i\frac{\pi}{2}\alpha_{0}^{2}}}{q^{3/4}\vartheta_{3}(0,q)4\alpha_{0}\beta_{0}\sqrt{\Upsilon_{M}(-2\alpha_{0},h)}}\label{eq5.29}
\eeq
We obtain:
\bea
<<\Phi_{c}(\infty)\Phi_{b}(1)\Phi_{a}(0)>>_{continued}\nn\\
=\sqrt{\Upsilon_{M}(-i(-2\beta_{0}),-ih)}\times [\Upsilon_{M}(-i(a+b+c-2\beta_{0}),-ih)\nn\\ 
\times\Upsilon_{M}(-i(-a+b+c),-ih)\Upsilon_{M}(-i(a-b+c),-ih)\Upsilon_{M}(-i(a+b-c),-ih)]\nn\\
/[\Upsilon_{M}(-2ia,-ih)\Upsilon_{M}(-i(2a-2\beta_{0}),-ih)
\Upsilon_{M}(-i2b,-ih)\nn\\ \times \Upsilon_{M}(-i(2b-2\beta_{0}),-ih)
\Upsilon_{M}(-i2c,-ih)\Upsilon_{M}(-i(2c-2\beta_{0}),-ih)]^{1/2}\nn\\
=
\frac{e^{i\frac{\pi}{2}\alpha_{0}^{2}}}{q^{3/4}\vartheta_{3}(0,q)4\alpha_{0}\beta_{0}\sqrt{\Upsilon_{M}(-2\alpha_{0},h)}}\nn\\
\frac{\sqrt{\Upsilon_{L}(2a,h)\Upsilon_{L}(2a-2\beta_{0},h)\Upsilon_{L}(2b,h)
\Upsilon_{L}(2b-2\beta_{0},h)\Upsilon_{L}(2c,h)\Upsilon_{L}(2c-2\beta_{0},h)}}
{\sqrt{\vartheta_{1}(\pi h2a)\vartheta_{1}(\pi h(2a-2\beta_{0}))\vartheta_{1}(\pi h2b)
\vartheta_{1}(\pi h(2b-2\beta_{0}))\vartheta_{1}(\pi h2c)
\vartheta_{1}(\pi h(2c-2\beta_{0}))}}\nn\\
\times\frac{\vartheta_{1}(\pi h(a+b+c-2\beta_{0}))\vartheta_{1}(\pi h(-a+b+c))
\vartheta_{1}(\pi h(a-b+c))\vartheta_{1}(\pi h(a+b-c))}
{\Upsilon_{L}(a+b+c-2\beta_{0},h)\Upsilon_{L}(-a+b+c,h)\Upsilon_{L}(a-b+c,h)
\Upsilon_{L}(a+b-c,h)}\label{eq5.30}
\eea

Now we shall rewrite the equation (\ref{eq5.30}) as follows:
\bea
<<\Psi_{c}(\infty)\Psi_{b}(1)\Psi_{a}(0)>>_{continued}\nn\\
/[\vartheta_{1}(\pi h(a+b+c-2\beta_{0}),q)\vartheta_{1}(\pi h(-a+b+c)<q)\nn\\
\times\vartheta_{1}(\pi h(a-b+c),h)\vartheta_{1}(\pi h(a+b-c),h)]\nn\\
\times[\vartheta_{1}(\pi h2a,q)\vartheta_{1}(\pi h(2a-2\beta_{0}),q)\vartheta_{1}(\pi h2b,q)\vartheta_{1}(\pi h(2b-2\beta_{0}),q)\nn\\
\times\vartheta_{1}(\pi h2c,q)\vartheta_{1}(\pi h(2c-2\beta_{0}),q)]^{1/2}\nn\\
=\sqrt{\Upsilon_{M}(-i(-2\beta_{0}),-ih)} [\Upsilon_{M}(-i(a+b+c-2\alpha_{0})\Upsilon_{M}(-i(-a+b+c),-ih)\nn\\
\times\Upsilon_{M}(-i(a-b+c),-ih)\Upsilon_{M}(-i(a+b-c),-ih)]\nn\\
/[\vartheta_{1}(\pi h(a+b+c-2\beta),q)\vartheta_{1}(\pi h(-a+b+c),q)\nn\\
\times\vartheta_{1}(\pi h(a-b+c),q)\vartheta_{1}(\pi h(a+b-c),q)]\nn\\
\times[\vartheta_{1}(\pi h2a,q)\vartheta_{1}(\pi h(2a-2b_{0}),q)\vartheta_{1}(\pi h2b,q)\vartheta_{1}(\pi h(2b-2\beta_{0}),q)\nn\\
\times\vartheta_{1}(\pi h2c,q)\vartheta_{1}(\pi h(2c-2\beta_{0}),q)]^{1/2}\nn\\
/[\Upsilon_{M}(-i2a,-ih)\Upsilon_{M}(-i(2a-2\beta_{0}),-ih)\Upsilon_{M}(-i2b,-ih)\Upsilon_{M}(-i(2b-2\beta_{0}),-ih)\nn\\
\times\Upsilon_{M}(-i2c,-ih)\Upsilon_{M}(-i(2c-2\beta_{0}),-ih)]^{1/2}\nn\\
=\frac{e^{i\frac{\pi}{2}\alpha^{2}_{0}}}{q^{3/4}\vartheta_{3}(0,q)4\alpha_{0}\beta_{0}\sqrt{\Upsilon_{M}(-2\alpha_{0},h)}}\nn\\
\times\frac{[\Upsilon_{L}(2a,h)\Upsilon_{L}(2a-2\beta_{0},h)\Upsilon_{L}(2b,h)\Upsilon_{L}(2b-2\beta_{0},h)\Upsilon_{L}(2c,h)\Upsilon_{L}(2c-2\beta_{0}h)]^{1/2}}
{\Upsilon_{L}(a+b+c-2\beta_{0},h)\Upsilon_{L}(-a+b+c,h)\Upsilon_{L}(a-b+c,h)\Upsilon_{L}(a+b-c,h)}\label{eq5.31}
\eea

The idea to organise, in this way, the analytically continued expression 
for \\ $<<\Phi_{c}\Phi_{b}\Phi_{c}>>$ is the following.

Passing from the 3-point function, for statistical models, to the 3-point function 
of Liouville, is delicate. The final test, or the definition, for the 3-point functions is, in fact, given by the 4 point functions, by their decomposition, or factorisation, into a product of two 3-point functions, with the sum over the states in the intermediate channel. 
As the spectrum, of the intermediate channel, is discrete, in the minimal model
(and equally in the generalised minimal model), while the spectrum of the Liouville theory is expected to be continuous [13,14], to pass from the sum, over the intermediate states, 
to the integral, in the decomposition of the 4-point functions, could be  organised by representing the initial sum as a sum over the residues, by adding an appropriate function which produce poles. And then  the sum of the residues could be expressed by the appropriate integral.

The role of $\vartheta_{1}$ functions in the denominator of (\ref{eq5.31}), in the l.h.s.,
might be that of providing the necessary poles.

Saying it differently, the appearance of these $\vartheta_{1}$ functions, the ones 
involving interactions (like $\vartheta_{1}(\pi h(a+b+c-2\beta_{0}),q)$ etc.), in the analytic continuation from minimal models to Liouville, might be interpreted as a sign, 
or a proof, that in fact the intermediate states spectrum of the Liouville is going to be continuous.

We are, actually, starting talking of the possibility to obtain the 4 point functions of Liouville by the analytic continuation 
of the well defined 4 point functions 
of minimal models. If realised, the associativity, in particular, will
not need to be proved, will be automatic.  
 
The task should be more complicated than that of continuing the 3-point functions. The possibility is to be attempted. For the moment we haven't yet much progressed in that direction.

The appearance of a product of "local" $\vartheta_{1}$ functions, under the square root, 
in the l.h.s of (\ref{eq5.31}), like $\vartheta_{1}(\pi h2a,q)$, $\vartheta(\pi h(2a-2\beta_{0}),q)$ etc., is related to the question of the appropriate normalisation, of the individual operators. We have putted them to the l.h.s., in (\ref{eq5.31}), so that they complete the product of $\Upsilon_{M}(-i2a,-ih)$ etc., which are also the normalisation factors, analytically continued. In particular in this way the common zeros, 
of $\Upsilon_{M}(-i2a,-ih)$ and $\vartheta_{1}(\pi h2a,q)$, etc., will be cancelled. But we would not insist on this point for the moment. It is the question of the appropriate normalisation of the Liouville vertex operators. The factor in front, in the r.h.s. of (\ref{eq5.31}), is also related to the question of  normalisation, of the 3-point function of Liouville. Might also to be decided by the proper definition of the 4-point function.

\numberwithin{equation}{section}

\vskip1.5cm

\section{Discussions.}

Historically, the first step of the analytical continuation, from the $(1,n)$ operators 3-point functions to the general $(n',n)$ minimal model 
operators 3-point functions, the continuation presented in the Section 2, 
it was realised long ago [15], as a by product of the curiosity, during my work on 3-point amplitudes of minimal models coupled to gravity [16].
At that time I have also defined the "naive" norms of vertex operators, 
$V_{a}(z,\bar{z})$ and $V_{a}^{+}(z,\bar{z})$, the analyses described in [11],
Section 9.1. This last curiosity was better justified, because in 3-point amplitudes of minimal models coupled to gravity, after cancellations, remain only products of norms of the operators.

At the end of the previous section we have started arguing that the final precisions for the definition of 3-point functions should be given by the 4-point ones, 
in which the 3-point functions participate "dynamically", in the sum over the states in the intermediate channel.

For instance, the question was raised in [6] with respect to apparent non-decoupling of some states from the outside of the minimal model (finite) 
Kac table of primary operators (the actual minimal model, 
not the generalised one). The problem that the 3-point functions (or operator algebra constants) with particular operators from outside the Kac table, 
do not vanish, the way they are defined analytically by the direct calculation of the 3-point functions.
This breaks the "fusion rules" of minimal models. So that some decouplings have to be added by hand.

We have seen the answer to this question in our work of  [8 - 10], 
where the operator algebra constants (3-point functions) have been derived from the structure of the 4-point functions of minimal models, which were the principal objects of [8 - 10]. 
We have seen that in the sum over the intermediate states, in the case of actual minimal models, 
it happens that for a particular the primary operator, which is placed 
outside of the Kac table, its contribution to the sum over the intermediate states gets cancelled by the contribution of a descendent operator 
of the nearby channel of another primary operator, positioned 
inside the table. 
And in this way the "fusion rules" get restored, analytically, not by hand.
One example of such "delicate decoupling", which could be seen only on the level of 4-point functions, is described in Section 9.2 of [11].

Above mentioned is just an example. But in general, we wish to stress again that, most likely, the proper, definite definition of 3-point functions should be provided by the 4-point ones. 
In particular, for the Liouville model.

\vskip1.5cm

{\bf Acknowledgments.}

I am grateful to Marco Picco and Raoul Santachiara for numerous 
useful discussions.

\vskip1.5cm

\appendix

\section{ Formulas needed for the analytic continuation 
of $C^{p}_{n,m}(\rho)$ in the Section 2.}

 In Section 2 we have defined the factors $g_{k}(\rho)$, $G_{lk}(\rho)$, $g^{(\alpha)}_{k}(\rho)$, $G^{(\alpha)}_{lk}(\rho)$, in (\ref{eq25}) - (\ref{eq28}), so that the functions $C^{p}_{n,m}(\rho)$ and $C^{(p',p)}_{(n',n)(m',m)}$ were given by the products in (\ref{eq29}) and (\ref{eq30}). To prove the formula (\ref{eq34}), 
 the first step of the analytic continuation, we have used the formulas (\ref{eq31}), (\ref{eq32}). We shall prove these relations now.
 
 \underline{$g_{k}(\rho)$, $G_{lk}(\rho)$.}
 
 Taking $\log$ of (\ref{eq25}) we get:
 \beq
 \log g_{k}(\rho)=\sum^{k}_{j=1}(\log\Gamma(j\rho)
 -\log\Gamma(1-j\rho))\label{eqA.1}
 \eeq
 Next we use the integral representation of $\Gamma(x)$:
 \beq
 \log\Gamma(x)=\int_{0}^{\infty}\frac{dt}{t}[(x-1)e^{-t}-\frac{e^{-t}-e^{-xt}}{1-e^{-t}}]\label{eqA.2}
 \eeq
  \beq
 \log\Gamma(x)-\log\Gamma(1-x)=\int_{0}^{\infty}\frac{dt}{t}[(2x-1)e^{-t}+\frac{e^{-xt}-e^{-(1-x)t}}{1-e^{-t}}]\label{eqA.3}
 \eeq
 One finds:
 \bea
 \log g_{k}(\rho)=\int_{0}^{\infty}\frac{dt}{t}\sum^{k}_{j=1}[(2j\rho-1)e^{-t}+\frac{e^{-j\rho t}-e^{-(1-j\rho)t}}{1-e^{-t}}]\nn\\
 =\int^{\infty}_{0}\frac{dt}{t}[(k(k+1)\rho-k)e^{-t}\nn\\
 +\frac{1}{1-e^{-t}}(\frac{e^{-\rho t}(1-e^{-k\rho t})}{1-e^{-\rho t}}-
 \frac{e^{-t}e^{\rho t}(1-e^{k\rho t})}{1-e^{\rho t}})]\label{eqA.4}
 \eea
 \bea
 \log g_{k}(\rho)=\int^{\infty}_{0}\frac{dt}{t}[(k(k+1)\rho-k)e^{-t}\nn\\
 +\frac{1}{1-e^{-t}}(\frac{e^{-\rho t}(1-e^{-k\rho t})}{1-e^{-\rho t}}
 +\frac{e^{-t}(1-e^{k\rho t})}{1-e^{-\rho t}})]\label{eqA.5}
\eea
In a similar way, by taking $\log$ of (\ref{eq26}) we find:
\bea
\log G_{lk}(\rho)=\int_{0}^{\infty}\frac{dt}{t}[(l(l+1)\rho'-2kl-l+k(k+1)\rho-k)e^{-t}\nn\\
+\int_{0}^{\infty}\frac{dt}{t}\frac{1}{(1-e^{-t})
(1-e^{-\rho't})}(e^{kt-\rho't}(1-e^{-l\rho't})+e^{-(1+k)t}(1-e^{l\rho't}))\nn\\
+\int^{\infty}_{0}\frac{dt}{t}\frac{1}{(1-e^{-t})
(1-e^{-\rho t})}(e^{-\rho t}(1-e^{-k\rho t})+e^{-t}(1-e^{k\rho t}))]\label{eqA.6}
\eea
Next, replacing $k$ by $k-\rho'l$ in (\ref{eqA.5}), we get:
\bea
\log g_{k-\rho'l}(\rho)=\int_{0}^{\infty}\frac{dt}{t}[((k-\rho'l)(k-\rho'l+1)\rho-(k-\rho'l))e^{-t}\nn\\
+\frac{1}{(1-e^{-t})(1-e^{-\rho t})}(e^{-\rho t}(1-e^{-(k-\rho'l)\rho t})
+e^{-t}(1-e^{(k-\rho'l)\rho t}))]\label{eqA.7}
\eea
Our purpose is to compare $\log G_{lk(\rho)}$, (\ref{eqA.6}), and $\log g_{k-\rho'l}(\rho)$, the expression above.

For the polynomial part, the coefficient of $e^{-t}
$ in (\ref{eqA.7}), one finds:
\bea
(k-\rho'l)(k-\rho'l+1)\rho - (k-\rho'l)\nn\\
=k(k+1)\rho-l(2k+1)+\rho'l^{2}-k+\rho'l\nn\\
=k(k+1)\rho - 2kl - l + \rho'l(l+1)-k\label{eqA.8}
\eea
This agrees with the polynomial part in (\ref{eqA.6}).

To give the exponential parts of (\ref{eqA.6}) and (\ref{eqA.7}) 
a similar appearance, we shall change the variable of integration $t$ in the first integral of the exponential part in (\ref{eqA.6}) as $t\rightarrow \rho t$, i.e. $t=\rho\tilde{t}$, and shall we drop the tilde afterwards.

We ignore for the moment the fact that the integral, of the exponential part taken alone, is divergent at $t\rightarrow 0$. We shall take care of the extra terms, the "anomaly" terms, which are due to this divergence,
a little bit later.

After the change $t\rightarrow \rho t$ in the first integral of the exponential part in (\ref{eqA.6}), the whole exponential part of (\ref{eqA.6}) takes the form:
\bea
\int_{0}^{\infty}\frac{dt}{t}\frac{1}{(1-e^{-\rho t}(1-e^{-t})}
[e^{k\rho t-t}(1-e^{-lt})+e^{-(1+k)\rho t}(1-e^{lt})]\nn\\
+\int_{0}^{\infty}\frac{dt}{t}\frac{1}{(1-e^{-t})
(1-e^{-\rho t})}[e^{-\rho t}(1-e^{-k\rho t})+e^{-t}(1-e^{k\rho t})]
+\mbox{anomaly}\nn\\
=\int_{0}^{\infty}\frac{dt}{t}\frac{1}{(1-e^{-t})(1-e^{-\rho t})}[e^{k\rho t-t}-e^{k\rho t-t-lt}+e^{-(1+k)\rho t}-e^{-(1+k)\rho t+lt}\nn\\
+e^{-\rho t}-e^{-\rho t-k\rho t}+e^{-t}-e^{-t+k\rho t}]
+\mbox{anomaly}\nn\\
=\int_{0}^{\infty}\frac{dt}{t}\frac{1}{(1-e^{-t})(1-e^{-\rho t})}
[e^{-\rho t}(1-e^{-k\rho t+lt})
+e^{-t}(1-e^{k\rho t-lt})]+\mbox{anomaly} \label{eqA.9}
\eea
Apart from the "anomaly", the expression above agrees with the exponential part of (\ref{eqA.7}). The agreement for the polynomial parts of (\ref{eqA.6}) and (\ref{eqA.7}) has been established earlier, (\ref{eqA.8}). So that we find the relation:
\beq
\log G_{lk}(\rho)=\log g_{k-\rho'l}(\rho)+\mbox{anomaly}\label{eqA.10}
\eeq

\vskip0.5cm

\underline{Anomaly.}

With respect to the divergence at $t\rightarrow 0$, the integral in (\ref{eqA.6}) for which we have changed the variable of integration, $t\rightarrow\rho t$, this integral should have been taken, more properly, in the form:
\beq
\lim_{\epsilon\rightarrow 0}\{\int_{\epsilon}^{\infty}\frac{dt}{t}\frac{1}{ (1-e^{-t})(1-e^{-\rho' t})}\nn\\
\times[e^{kt-\rho't}(1-e^{-l\rho't})+e^{-(1+k)t}(1-e^{l\rho't})]\}\label{eqA.11a}
\eeq
After the change $t\rightarrow\rho t$, ($t=\rho\tilde{t}$, and we drop the tilde afterwards) we obtain:
\bea
\lim_{\epsilon\rightarrow 0}\{\int^{\infty}_{\frac{\epsilon}{\rho}}\frac{dt}{t}\frac{1}{(1-e^{-\rho t})(1-e^{-t})}
\times[e^{\rho kt-t}(1-e^{-lt})+e^{-(1+k)\rho t}(1-e^{lt})]\}\nn\\
=\lim_{\epsilon\rightarrow 0}\{\int_{\epsilon}^{\infty}\frac{dt}{t}\frac{1}{(1-e^{-\rho t})(1-e^{-t})}
\times[e^{\rho kt-t}(1-e^{-lt})+e^{-(1+k)\rho t}(1-e^{lt})]\nn\\
+\int^{\epsilon}_{\frac{\epsilon}{\rho}}\frac{dt}{t}\frac{1}{(1-e^{-\rho t})(1-e^{-t})}
\times[e^{\rho kt-t}(1-e^{-lt})+e^{-(1+k)\rho t}(1-e^{lt})]\}\label{eqA.11}
\eea
The first integral in the expression above goes to joint the rest, the polynomial part and the second integral of the exponential part in (\ref{eqA.6}), which also should have been treated properly with the limit $\epsilon \rightarrow 0$ for the lower limit of integration. 

But the second integral in (\ref{eqA.11}) above gives the additional term, which we have called  "anomaly". We find (the limit $\epsilon \rightarrow 0 $ is assumed):
\bea
\mbox{anomaly} = \int^{\epsilon}_{\frac{\epsilon}{\rho}}\frac{dt}{t}\frac{1}{\rho t^{2}}
[(1+\rho kt-t)(lt-\frac{1}{2}l^{2}t^{2})
+(1-(1+k)\rho t)(-lt-\frac{1}{2}l^{2}t^{2})]\nn\\
=\int^{\epsilon}_{\frac{\epsilon}{\rho}}\frac{dt}{t}\frac{1}{\rho t^{2}}[(2kl+l)\rho t^{2}-l(l+1)t^{2}]
=\int^{\epsilon}_{\frac{\epsilon}{\rho}}\frac{dt}{t}[2kl+l-\rho'l(l+1)]\nn\\
=\log\rho\cdot(2kl+l-\rho'l(l+1))\label{eqA.12}
\eea
Finaly, we get, by (\ref{eqA.10}) and (\ref{eqA.12}),
\beq
\log G_{k,l}(\rho)=\log g_{k-\rho'l}(\rho)+\log\rho\cdot(2kl+l-\rho'l(l+1))\label{eqA.13}
\eeq
which is the relation (\ref{eq31}).

The second relation, eq.(\ref{eq32}), is derived in the same way, by starting with the $\log$'s of products for $g^{(\alpha)}_{k}(\rho)$ and $G_{lk}^{(\alpha)}(\rho)$ in (\ref{eq27}) 
and (\ref{eq28}).

\vskip0.8cm

\underline{{\bf Translation relations of $\Upsilon_{M}(x,h)$.}}

Just for completeness, we shall show how the discrete translation relations 
for the function $\Upsilon_{M}(x,h)$ [5,6], the relations in (\ref{eqB.1}), (\ref{eqB.2}), could be derived.

We shall show it with the relation (\ref{eqB.1}). We get: 
\bea
\log\Upsilon_{M}(x+h)=\int_{0}^{\infty}\frac{dt}{t}\{(\alpha_{0}-x-h)^{2}e^{-t}-\frac{\sinh^{2}[(\alpha_{0}-x-h)\frac{t}{2}]}{\sinh\frac{th}{2}\cdot\sinh\frac{t}{2h}}\}\nn\\
=\int_{0}^{\infty}\frac{dt}{t}\{[(\alpha_{0}-x)^{2}-2h(\alpha_{0}-x)+h^{2}]e^{-t}-\frac{\sinh^{2}[(\alpha_{0}-x-h)\frac{t}{2}]}{\sinh\frac{th}{2}\cdot\sinh\frac{t}{2h}}\}\label{eqA.14a}
\eea
As $\alpha_{0}=\frac{h}{2}-\frac{1}{2h}$, one gets:
\beq
\log\Upsilon_{M}(x+h)=\int_{0}^{\infty}\frac{dt}{t}\{[(\alpha_{0}-x)^{2}+2hx+1]e^{-t}-\frac{\sinh^{2}[(\alpha_{0}-x-h)\frac{t}{2}]}{\sinh\frac{th}{2}\cdot\sinh\frac{t}{2h}}\}\label{eqA.14}
\eeq
We consider
\bea
\log\gamma(-hx)=\log\Gamma(-hx)-\log\Gamma(1+hx)\nn\\
=\int_{0}^{\infty}\frac{dt}{t}\{(-2hx-1)e^{-t}+\frac{e^{hxt}-e^{-(1+hx)t}}{1-e^{-t}}\}\label{eqA.15}
\eea
-- to compensate the extra terms in the polynomial part of (\ref{eqA.14}). We have used the integral representation in (\ref{eqA.3}).

For the sum of (\ref{eqA.14}) and (\ref{eqA.15}) we obtain:
\bea
\log\Upsilon_{M}(x+h)+\log\gamma(-hx)\nn\\
=\int_{0}^{\infty}\frac{dt}{t}\{(\alpha_{0}-x)^{2} e^{-t}+\frac{e^{hxt}-e^{-(1+hx)t}}{1-e^{-t}}-\frac{\sinh^{2}[(\alpha_{0}-x-h)\frac{t}{2}]}{\sinh\frac{th}{2}\cdot\sinh\frac{t}{2h}}\}\label{eqA.16}
\eea
The first term in the integral above is already that of
\beq
\log\Upsilon_{M}(x,h)=\int_{0}^{\infty}\frac{dt}{t}\{(\alpha_{0}-x)^{2}e^{-t}
-\frac{\sinh^{2}[(\alpha_{0}-x)\frac{t}{2}]}{\sinh\frac{th}{2}\cdot\sinh\frac{t}{2h}}\}\label{eqA.17}
\eeq
We shall transform, separately, the second and the third terms in (\ref{eqA.16}).

By transforming $t\rightarrow\frac{t}{h}$ in the integral of the second term in (\ref{eqA.16}), we obtain:
\beq
\int_{0}^{\infty}\frac{dt}{t}\frac{e^{xt}-e^{-\frac{t}{h}-xt}}{1-e^{-\frac{t}{h}}}\label{eqA.18}
\eeq
and we continue its transformation as follows:
\bea
=\int_{0}^{\infty}\frac{dt}{t}\frac{(e^{xt}-e^{-\frac{t}{h}-xt})(1-e^{-ht})}{(1-e^{-\frac{t}{h}})(1-e^{-ht})}
=\int^{\infty}_{0}\frac{dt}{t}\frac{e^{\frac{ht}{2}+\frac{t}{2h}}(e^{xt}-e^{-\frac{t}{h}-xt})(1-e^{-ht})}{4\sinh\frac{t}{2h}\cdot\sinh\frac{ht}{2}}\nn\\
=\int^{\infty}_{0}\frac{dt}{t} \frac{e^{\frac{ht}{2}+\frac{t}{2h}+xt}-e^{\frac{ht}{2}-\frac{t}{2h}-xt}-e^{-\frac{ht}{2}+\frac{t}{2h}+xt}+e^{-\frac{ht}{2}-\frac{t}{2h}-xt}}{4\sinh\frac{t}{2h}\cdot\sinh\frac{hx}{2}}\label{eqA.19}
\eea
Now we shall transform, a little bit, the third term in the integral of (\ref{eqA.16}):
\bea
\int^{\infty}_{0}\frac{dt}{t}\{-\frac{e^{(\alpha_{0}-x-h)t}
+e^{-(\alpha_{0}-x-h)t}-2}{4\sinh\frac{th}{2}\cdot\sinh\frac{t}{2h}}\}\nn\\
=\int^{\infty}_{0}\frac{dt}{t}\{-\frac{e^{-\frac{ht}{2}-\frac{t}{2h}-xt}+e^{\frac{ht}{2}
+\frac{t}{2h}+xt}-2}{4\sinh\frac{th}{2}\cdot\sinh\frac{t}{2h}}\}\label{eqA.20}
\eea
For the sum of (\ref{eqA.19}) and (\ref{eqA.20}), which is the sum of the second and the third terms in (\ref{eqA.16}), we obtain:
\bea
\int^{\infty}_{0}\frac{dt}{t}
\{-\frac{e^{\frac{ht}{2}-\frac{t}{2h}-xt}+e^{-\frac{ht}{2}+\frac{t}{2h}+xt}-2}{4 \sinh\frac{th}{2}\cdot\sinh\frac{t}{2h}}\}\nn\\
=-\int^{\infty}_{0}\frac{dt}{t}
\frac{e^{\alpha_{0}t-xt}+e^{-\alpha_{0}t+xt}-2}{4 \sinh\frac{th}{2}\cdot\sinh\frac{t}{2h}}\nn\\
=-\int^{\infty}_{0}\frac{dt}{t}
\frac{\sinh^{2}[(\alpha_{0}-x)\frac{t}{2}]}{\sinh\frac{th}{2}\cdot\sinh\frac{t}{2h}}\label{eqA.21}
\eea
Together with he first term in (\ref{eqA.16}) this gives
\beq
\log\Upsilon_{M}(x+h)+\log\gamma(-hx)=\log\Upsilon_{M}(x)+\mbox{anomaly}\label{eqA.22}
\eeq
Here "anomaly" stands for the term which we have disregarded so far, 
by ignoring the divergence, at $t\rightarrow 0$, of the second integral in(\ref{eqA.16}), 
while transforming $t\rightarrow\frac{t}{h}$.

Properly, we should have defined it as
\beq
\lim_{\epsilon\rightarrow 0}\{\int_{\epsilon}^{\infty}\frac{dt}{t}\frac{e^{hxt}-e^{-(1+hx)t}}{1-e^{-t}}\}\label{eqA.23}
\eeq
With $t=\frac{\tilde{t}}{h}$, we obtain:
\beq
\lim_{\epsilon\rightarrow 0}\{\int_{h\epsilon}^{\infty}\frac{d\tilde{t}}{\tilde{t}}\frac{e^{x\tilde{t}}-e^{-\frac{\tilde{t}}{h}-x\tilde{t}}}{1-e^{-\frac{\tilde{t}}{h}}}\}\label{eqA.24}
\eeq
We drop the tildes and we rewrite (\ref{eqA.24}) as follows:
\beq
\lim_{\epsilon\rightarrow 0}\{\int_{\epsilon}^{\infty}\frac{dt}{t}
\frac{e^{xt}-e^{-\frac{t}{h}-xt}}{1-e^{-\frac{t}{h}}}+\int_{h\epsilon}^{\epsilon}\frac{dt}{t}\frac{e^{xt}-e^{-\frac{t}{h}-xt}}{1-e^{-\frac{t}{h}}}\label{eqA.25}
\eeq
The first integral goes to join our derivation of the relation (\ref{eqA.22}), with the exception of the anomaly in it, while the second integral in (\ref{eqA.25}) gives the anomaly.

We get:
\bea
\mbox{anomaly}=\lim_{\epsilon\rightarrow 0}\{\int_{h\epsilon}^{\epsilon}\frac{dt}{t}\frac{e^{xt}-e^{-\frac{t}{h}-xt}}{1-e^{-\frac{t}{h}}}\}
=\lim_{\epsilon\rightarrow 0}\{\int_{h\epsilon}^{\epsilon}\frac{dt}{t}\frac{xt+\frac{t}{h}+xt}{\frac{t}{h}}\}\nn\\
=\lim_{\epsilon\rightarrow 0}\{\int_{h\epsilon}^{\epsilon}\frac{dt}{t}(2xh+1)\}
=-(2xh+1)\cdot\log h  \label{eqA.26}
\eea
Finally, the relation (\ref{eqA.22}) takes the form:
\beq
\log\Upsilon_{M}(x+h)=-\log\gamma(-hx)+\log\Upsilon_{M}(x)-(2xh+1)\log h\label{eqA.27}
\eeq
\beq
\Upsilon_{M}(x+h)=\frac{1}{\gamma(-hx)}h^{-2xh-1}\Upsilon_{M}(x)\label{eqA.28}
\eeq
This is the relation (\ref{eqB.1}).

\vskip1.5cm

\section{ The function $\Upsilon_{M}(x,h)$ and the list of some of its values. The list of results for particular 3-point functions which are used in the Section 4.}

We shall remind here certain properties of the function $\Upsilon_{M}(x)$ [5,6]. We shall list some particular values of this function. 
And we shall also derive several expressions for the 3-point functions which are used in the Section 4.

The function $\Upsilon_{M}(x)\equiv\Upsilon_{M}(x,h)$, defined in (\ref{eq3.28}), satisfies the following "quasi-periodicity" relations, with respect to translations by $h$ and by $1/h$:
\beq
\Upsilon_{M}(x+h)=\frac{1}{\gamma(-hx)}h^{-1-2hx}\times\Upsilon_{M}(x)\label{eqB.1}
\eeq
$\gamma(x)=\Gamma(x)/\Gamma(1-x)$.
\beq
\Upsilon_{M}(x-\frac{1}{h})=\frac{1}{\gamma(\frac{x}{h})}h^{1-\frac{2x}{h}}\times\Upsilon_{M}(x)\label{eqB.2}
\eeq
\beq
\Upsilon_{M}(x)=\gamma(-hx)h^{1+2hx}\times\Upsilon_{M}(x+h)\label{eqB.3}
\eeq
\beq
\Upsilon_{M}(x)=\gamma(\frac{x}{h})h^{-1+\frac{2x}{h}}\times\Upsilon_{M}(x-\frac{1}{h})\label{B.4}
\eeq
\beq
\Upsilon_{M}(x-h)=\gamma(-h(x-h))h^{1+2h(x-h)}\times\Upsilon_{M}(x)\label{eqB.5}
\eeq
\beq
\Upsilon_{M}(x+\frac{1}{h})=\gamma(\frac{1}{h}(x+\frac{1}{h}))h^{-1+\frac{2}{h}(x+\frac{1}{h})}\times\Upsilon_{M}(x)\label{eqB.6}
\eeq
We have listed these properties in various forms, which are useful in actual calculations, though, evidently, (\ref{eqB.3}) - (\ref{eqB.6}) follow directly from (\ref{eqB.1}), (\ref{eqB.2}). 
The derivation of (\ref{eqB.1}) is given in the Appendix A. 

The obvious property, by (\ref{eq3.28}), is that
\beq
\Upsilon_{M}(2\alpha_{0}-x)=\Upsilon_{M}(x)\label{eqB.7}
\eeq

Next, one finds directly, from the integral form of $\log\Upsilon_{M}(x)$ in (\ref{eq3.28}), that

1.
\bea
\Upsilon_{M}(x)\simeq x+\frac{1}{h},\quad x\rightarrow -\frac{1}{h}\nn\\
\Upsilon_{M}(-\frac{1}{h})=0\label{eqB.8}
\eea

2.
\bea
\Upsilon_{M}(x)\simeq h-x,\quad x\rightarrow h\nn\\
\Upsilon_{M}(h)=0\label{eqB.9}
\eea
The integral in (\ref{eq3.28}), which defines $\Upsilon_{M}(x,h)$, is convergent 
for $-\frac{1}{h}<x<h$. Outside, the function is defined by the analytic continuation: by the translations (\ref{eqB.1}) - (\ref{eqB.6}). For $x\rightarrow-\frac{1}{h}$, from above, and for $x\rightarrow h$, from below, the integral in (\ref{eq3.28}) is logarithmically divergent at $t\rightarrow\infty$, which results, 
when evaluated, in the asymptotics in (\ref{eqB.8}), (\ref{eqB.9}). 

3. In general, the zeros of $\Upsilon_{M}(x)$, which we denote as $x^{(M)}_{n.m}$, are located at
\bea
x^{(M)}_{n,m}=-\frac{1}{h}(n+1)-hm,\nn\\
\mbox{and at} \quad x^{(M)}_{n,m}=\frac{1}{h}n+h(m+1)\label{eqB.10}
\eea
$n,m=0,1,2,3,...$

They could be obtained, with some care, from (\ref{eqB.8}), (\ref{eqB.9}) by the translations 
(\ref{eqB.1}) - (\ref{eqB.6}). 

4. Next, it is evident from (\ref{eq3.28}) that
\beq
\Upsilon_{M}(\alpha_{0})=1  \label{eqB.11}
\eeq

5. One also finds that
\beq
\Upsilon_{M}(0)=\Upsilon_{M}(2\alpha_{0})=1\label{eqB.12}
\eeq
In fact, by (\ref{eqB.5}) and (\ref{eqB.9}), for $x\rightarrow h$, one finds:
\bea
\Upsilon_{M}(x-h)=\gamma(-h(x-h))h^{1+2h(x-h)}\times\Upsilon_{M}(x)\nn\\
\simeq\frac{1}{-h(x-h)}h\times(h-x)=1,\nn\\
\Upsilon_{M}(0)=1\label{eqB.13}
\eea
$\Upsilon_{M}(2\alpha_{0})=1$ follow by the property (\ref{eqB.7}).

The next several values of $\Upsilon_{M}(x)$  are obtained by translations (\ref{eqB.1}) - (\ref{eqB.6}). One finds:

6.
\beq
\Upsilon_{M}(2h-\frac{1}{h})=\Upsilon_{M}(h+2\alpha_{0})=\Upsilon_{M}(-h)
=\gamma(h^{2})\cdot h^{1-2h^{2}}=\gamma(\rho)\cdot\rho^{\frac{1}{2}-\rho}\label{eqB.14}
\eeq

7.
\beq
\Upsilon_{M}(\frac{1}{h})=\gamma(\frac{1}{h^{2}})h^{\frac{2}{h^{2}}-1}=\gamma(\rho')\rho^{\rho'-\frac{1}{2}}\label{eqB.15}
\eeq

8.
\beq
\Upsilon_{M}(-h)=\gamma(\rho)\rho^{\frac{1}{2}-\rho}\label{eqB.16}
\eeq

9.
\bea
\Upsilon_{M}(-2\alpha_{0})=\Upsilon_{M}(-h+\frac{1}{h})\nn\\
=\gamma(\rho-1)\gamma(\rho')\rho^{1-\rho+\rho'}\nn\\
=-\frac{\rho}{(\rho-1)^{2}}\gamma(\rho)\gamma(\rho')\rho^{-\rho+\rho'}\label{eqB.17}
\eea

10.
\bea
\Upsilon_{M}(-2h+\frac{1}{h})=\Upsilon_{M}(-h-2\alpha_{0})\nn\\
=-\frac{1}{(1-\rho)^{2}}\gamma(2\rho-1)\gamma(\rho)\gamma(\rho')\rho^{\frac{5}{2}-3\rho+\rho'}\label{eqB.18}
\eea

11.
\bea
\Upsilon_{M}(-h+\frac{2}{h})=\Upsilon_{M}(\frac{1}{h}-2\alpha_{0})\nn\\
=-\frac{1}{(1-\rho')^{2}}\gamma(2\rho'-1)\gamma(\rho)\gamma(\rho')\rho^{-\frac{5}{2}-\rho+3\rho'}\label{eqB.19}
\eea

12.
\beq
\Upsilon_{M}(-4\alpha_{0})=\gamma(2\rho-1)\gamma(2\rho'-1)\rho^{1-2\rho+2\rho'}\times\Upsilon_{M}(-2\alpha_{0})\label{eqB.20}
\eeq

We shall derive now several results for the correlation functions, the results listed in the Section 4.

\underline{$<III>.$}

$I=V_{1,1}$. $a=b=c=0$. By (\ref{eq4.6}), $l=k=1$.

By the formula (\ref{eq4.9}),
\bea
<III>=\frac{\Upsilon_{M}(-2\alpha_{0})(\Upsilon_{M}(0))^{3}}{(\Upsilon_{M}(0))^{3}}\times\rho^{1-\rho'}\times(\rho')^{1-\rho}\nn\\
=\Upsilon_{M}(-2\alpha_{0})\times\rho^{\rho-\rho'}=-\frac{\rho}{(1-\rho)^{2}}\gamma(\rho)\gamma(\rho')\label{eqB.21}
\eea
-- if we use the value (\ref{eqB.17}) for $\Upsilon_{M}(-2\alpha_{0})$.

By the formula (\ref{eq4.8}), if we put $a=0$, $b=0$, $c=0$ directly, we shall get a problem, the expression will not be defined. But the integral, which gives (\ref{eq4.8}), is defined, in fact, with a single condition,
\beq 
a+b+c+l\alpha_{-}+k\alpha_{-}=2\alpha_{0}\label{eqB.22}
\eeq
with $l,k$ being integers. Separately, $a,b,c$ do not have to be degenerate, to make the expression (\ref{eq4.8}) valid. This allows to define the function $<III>$, with (\ref{eq4.8}), 
as a limit $a\rightarrow 0$, $b\rightarrow 0$, $c\rightarrow 0$, while keeping $a+b+c=0$ to make the condition (\ref{eqB.22}) satisfied, with $l=k=1$.

In this way, for $\alpha,\beta,\gamma,\alpha',\beta',\gamma'$ being small, and $l=k=1$, we get, by (\ref{eq4.8}):
\bea
<V_{c} V_{b} V_{a}>=\rho^{-4}\times\frac{1}{(\rho'-1)^{2}}\frac{\Gamma(\rho')}{\Gamma(1-\rho')}\frac{\Gamma(\rho)}{\Gamma(1-\rho)}\nn\\
\times\frac{1}{(\alpha')^{2}(\beta')^{2}(\gamma')^{2}}\times\frac{\Gamma(1+\alpha')\Gamma(1+\beta')\Gamma(1+\gamma')}{\Gamma(-\alpha')\Gamma(-\beta')\Gamma(-\gamma')}\times\frac{\Gamma(1+\alpha)\Gamma(1+\beta)\Gamma(1+\gamma)}{\Gamma(-\alpha)\Gamma(-\beta)\Gamma(-\gamma)}\nn\\
\simeq\rho^{-4}\frac{1}{(\rho'-1)^{2}} \gamma(\rho')\gamma(\rho)
\frac{1}{(\alpha')^{2}(\beta')^{2}(\gamma')^{2}}
(-\alpha')(-\beta')(-\gamma')\times(-\alpha)(-\beta)(-\gamma)\label{eqB.23}
\eea
Since $\alpha=-\rho\alpha'$, $\alpha'\alpha=-\rho(\alpha')^{2}$, etc., we get
\bea
<V_{c} V_{b} V_{a}>\simeq\frac{\rho^{-4}}{(\rho'-1)^{2}}\times\gamma(\rho')\gamma(\rho)\times(-\rho^{3})\nn\\
=-\frac{\rho^{-1}}{(1-\rho)^{2}\cdot(\rho')^{2}}\gamma(\rho')\gamma(\rho)=-\frac{\rho}{(1-\rho)^{2}}\gamma(\rho')\gamma(\rho)\label{eqB.25}
\eea
In the limite $a\rightarrow0$, $b\rightarrow0$, $c\rightarrow0$ we get
\beq
<V_{0} V_{0} V_{0}>=<III>=-\frac{\rho}{(1-\rho)^{2}}\gamma(\rho')\gamma(\rho)\label{eqB.26}
\eeq
which agrees with (\ref{eqB.21}).

\underline{$<I^{+}II>$.}

\beq
<I^{+}II>=<V_{2\alpha_{0}} V_{0} V_{0}>=<V_{-1,-1} V_{1,1} V_{1,1}>\label{eqB.27}
\eeq
By (\ref{eq4.6}), $l=k=0$. Then, by (\ref{eq4.8}), one gets $<I^{+}II>=1$ immediately, as
\beq
\prod^{0}_{i=1}(...)=\prod^{-1}_{i=0}(...)=1\label{eqB.28}
\eeq

By(\ref{eq4.9}),
\beq
<I^{+}II>=\frac{\Upsilon_{M}(0)\Upsilon_{M}(2\alpha_{0})\Upsilon_{M}(2\alpha_{0})\Upsilon(-2\alpha_{0})}{\Upsilon_{M}(0)\Upsilon_{M}(0)\Upsilon(4\alpha_{0})}\label{eqB.29}
\eeq
Since $\Upsilon(0)=\Upsilon(2\alpha_{0})=1$, $\Upsilon(4\alpha_{0})
=\Upsilon(-2\alpha_{0})$ by (\ref{eqB.7}), we get equally
\beq
<I^{+}II>=1\label{eqB.30}
\eeq

\underline{$<I^{+} V_{1,2} V_{1,2}>$.}

For this function $l=0$, $k=1$. By (\ref{eq4.8}):
\bea
<I^{+} V_{1,2} V_{1,2}>=\frac{\Gamma(\rho)}{\Gamma(1-\rho)}\times(\frac{\Gamma(1+\alpha)}{\Gamma(-\alpha)})^{2}\times\frac{\Gamma(1+\gamma)}{\Gamma(-\gamma)}\nn\\
\alpha=2 \alpha_{1,2} \alpha_{+} = 2(-\frac{\alpha_{+}}{2})\alpha_{+}=-\rho,\nn\\
\gamma=2\cdot2\alpha_{0}\cdot\alpha_{+}=2(\alpha_{+}+\alpha_{-})\alpha_{+}=2\rho-2,\nn\\
<I^{+} V_{1,2} V_{1,2}>=\frac{\Gamma(\rho)}{\Gamma(1-\rho)}\cdot(\frac{\Gamma(1-\rho)}{\Gamma(\rho)})^{2}\cdot\frac{\Gamma(2\rho-1)}{\Gamma(2-2\rho)}\nn\\
=\frac{\Gamma(1-\rho)}{\Gamma(\rho)}\cdot\frac{\Gamma(2\rho-1)}{\Gamma(2-2\rho)}=\frac{\gamma(2\rho-1)}{\gamma(\rho)}\label{eqB.31}
\eea
-- which agrees with (\ref{eq4.16}).

By (\ref{eq4.9}), $a=b=-\frac{\alpha_{+}}{2}$, $c=2\alpha_{0}$,
\bea
<I^{+} V_{1,2} V_{1,2}>
=\frac{\Upsilon_{M}(-\alpha_{+})\Upsilon_{M}(2\alpha_{0})\Upsilon_{M}2\alpha_{0})\Upsilon(-\alpha_{+}-2\alpha_{0})}{\Upsilon_{M}(-\alpha_{+})\Upsilon_{M}(-\alpha_{+})\Upsilon(4\alpha_{0})}\times(\rho')^{1-\rho}\nn\\
=\frac{\Upsilon_{M}(-\alpha_{+}-2\alpha_{0})}{\Upsilon_{M}(-\alpha_{+})\Upsilon_{M}(-2\alpha_{0})}\rho^{-1+\rho}=\frac{\Upsilon_{M}(-h-2\alpha_{0})}{\Upsilon_{M}(-h)\Upsilon_{M}(-2\alpha_{0})}\rho^{-1+\rho}\label{eqB.32}
\eea
By (\ref{eqB.5}),
\bea
\Upsilon_{M}(-h-2\alpha_{0})=\gamma(-h(-h-2\alpha_{0}))h^{1+2h(-h-2\alpha_{0})}\times\Upsilon_{M}(-2\alpha_{0})\nn\\
=\gamma(-h(-2h+\frac{1}{h}))h^{1+2h(-2h+\frac{1}{h})}\times\Upsilon_{M}(-2\alpha_{0})\nn\\
=\gamma(2\rho-1)h^{3-4\rho}\times\Upsilon_{M}(-2\alpha_{0})\label{eqB.33}
\eea
Putting it into (\ref{eqB.32}), we get
\beq
<I^{+} V_{1,2} V_{1,2}>=\frac{1}{\Upsilon_{M}(-h)}\times\gamma(2\rho-1)\cdot\rho^{\frac{3}{2}-2\rho}\times\rho^{-1+\rho}\label{eqB.34}
\eeq
We take now the value for $\Upsilon_{M}(-h)$ in (\ref{eqB.16}). We obtain:
\beq
<I^{+} V_{1,2} V_{1,2}>=\frac{1}{\gamma(\rho)}\rho^{-\frac{1}{2}+\rho}\times\gamma(2\rho-1)\rho^{\frac{3}{2}-2\rho}\times\rho^{-1+\rho}=\frac{\gamma(2\rho-1)}{\gamma(\rho)}\label{eqB.35}
\eeq
This agrees with (\ref{eqB.31}) and with (\ref{eq4.15}), (\ref{eq4.16}).

\underline{$<IV_{1,2}V_{1,2}>=<V_{1.1}V_{1,2}V_{1,2}>.$}

We shall calculate it with (\ref{eq4.9}), which is simpler.

In this case $a=b=-\frac{\alpha_{+}}{2}$, $c=0$; $l=1$, $k=2$.
\bea
<IV_{1,2}V_{1,2}>=\frac{\Upsilon_{M}(-\alpha_{+}-2\alpha_{0})\Upsilon_{M}(0)\Upsilon_{M}(0)\Upsilon_{M}(-\alpha_{+})}{\Upsilon_{M}(-\alpha_{+})\Upsilon_{M}(-\alpha_{+})\Upsilon_{M}(0)}
\rho^{1-\rho'}(\rho')^{2(1-\rho)}\nn\\
=\frac{\Upsilon_{M}(-\alpha_{+}-2\alpha_{0})}{\Upsilon_{M}(-\alpha_{+})}\rho^{1-\rho'}\rho^{-2+2\rho}
=\frac{\Upsilon_{M}(-\alpha_{+}-2\alpha_{0})\times\rho^{-1+\rho}}{\Upsilon_{M}(-\alpha_{+})
\times\Upsilon_{M}(-2\alpha_{0})}\Upsilon_{M}(-2\alpha_{0})\rho^{-\rho'+\rho}\label{eqB.36}
\eea
The first factor is $<I^{+}V_{1,2}V_{1,2}>=(N_{1,2})^{2}$, according to (\ref{eqB.32}), and the second factor is $Z=<III>$, according to (\ref{eqB.21}). So that we get
\beq
<IV_{1,2}V_{1,2}>=Z (N_{1,2})^{2}\label{eqB.37}
\eeq
which confirms (\ref{eq4.18}).

\underline{$<I^{+}V_{1,2}^{+}V_{1,2}>=<V_{-1,-1}V_{-1,-2}V_{1,2}>.$}

In this case $a=-\frac{\alpha_{+}}{2}$, $b=\frac{3}{2}\alpha_{+} + \alpha_{-} $, $c=2\alpha_{0}$; 
$l=(-1-1+1-1)/2=-1$, $k=(-1-2+2-1)/2=-1$. With $l=-1$, $k=-1$, the use of the formula (\ref{eq4.8}) 
is blocked. We shall calculate this function with (\ref{eq4.9}).
\bea
<I^{+}V_{1,2}^{+}V_{1,2}>
=\frac{\Upsilon_{M}(2\alpha_{0})\Upsilon_{M}(-\alpha_{+})\Upsilon_{M}(4\alpha_{0}+\alpha_{+})\Upsilon_{M}(0)}{\Upsilon_{M}(3\alpha_{+}+2\alpha_{-})\Upsilon_{M}(-\alpha_{+})\Upsilon_{M}(4\alpha_{0})}
\rho^{-(1-\rho')}(\rho')^{-(1-\rho)} \nn\\
=\frac{\Upsilon_{M}(3\alpha_{+}+2\alpha_{-})}{\Upsilon_{M}(3\alpha_{+}+2\alpha_{-})\Upsilon_{M}(-2\alpha_{0})}
\rho^{-1+\rho'}\rho^{1-\rho}=\frac{1}{\Upsilon_{M}(-2\alpha_{0})\rho^{-\rho'+\rho}}=\frac{1}{Z}\label{eqB.38}
\eea
-- according to (\ref{eqB.21}). This agrees with (\ref{eq4.20}).

We shall derive still one more result, which is claimed in (\ref{eq4.38}).

\underline{$<I^{+}V^{+}_{a}V^{+}_{a}>=<I^{+}V_{a^{+}}V_{a^{+}}>$.}

Here
\bea 
a^{+}=2\alpha_{0}-\alpha_{n'.n}=\frac{1+n'}{2}\alpha_{+}+\frac{1+n}{2}\alpha_{-}, \quad c=2\alpha_{0};\nn\\
<I^{+}V_{a^{+}}V_{a^{+}}>=<V_{-1,-1}V_{-n',-n}V_{-n',-n}>,\nn\\
l=\frac{-1-2n'-1}{2}=-n'-1, \quad k=-n-1
\eea
We get:
\bea
<I^{+}V_{a}^{+}V_{a}^{+}>
=\frac{\Upsilon_{M}(4\alpha_{0}-2a)\Upsilon_{M}(2\alpha_{0})\Upsilon_{M}(2\alpha_{0})\Upsilon_{M}(2\alpha_{0}-2a)}{\Upsilon_{M}(4\alpha_{0}-2a)\Upsilon_{M}(4\alpha_{0}-2a)\Upsilon_{M}(4\alpha_{0})}\nn\\
\times\rho^{-(n'+1)(1-\rho')}\times(\rho')^{-(n+1)(1-\rho)}
=\frac{\Upsilon_{M}(2a)}{\Upsilon_{M}(2a-2\alpha_{0})\Upsilon_{M}(-2\alpha_{0})}\nn\\
\times\rho^{-(n'-1)(1-\rho')}\times(\rho')^{-(n-1)(1-\rho)}\times\rho^{-2(1-\rho')}\times(\rho')^{-2(1-\rho)}\label{eqB.39}
\eea
We have used the property (\ref{eqB.7}) for $\Upsilon_{M}(2\alpha_{0}-a)$, $\Upsilon_{M}(4\alpha_{0}-2a)$ and $\Upsilon_{M}(4\alpha_{0})$. We find:
\bea
<I^{+}V^{+}_{a}V^{+}_{a}>=\frac{\Upsilon_{M}(2a)\Upsilon_{M}(-2\alpha_{0})}{\Upsilon_{M}(2a-2\alpha_{0})}
\times\rho^{-(n'-1)(1-\rho')}\times(\rho')^{-(n-1)(1-\rho)}\nn\\
\times\frac{1}{(\Upsilon_{M}(-2\alpha_{0})\rho^{-\rho'+\rho})^{2}}=\frac{1}{(N_{a})^{2}\times Z^{2}}\label{eqB.40}
\eea
--according to (\ref{eq4.37}) and (\ref{eqB.21}). This result agrees with (\ref{eq4.38}).

\vskip1.5cm

\section{ Series for $\log[\vartheta_{1}(u)/\vartheta_{3}(0)]$.}

$\vartheta$ functions, $\vartheta_{1}(u)$ and $\vartheta_{3}(u)$, could be represented by infinite products [12]:
\bea
\vartheta_{1}(u)=2q^{1/4}\cdot\sin u\cdot\prod^{\infty}_{m=1}(1-2q^{2m}\cdot\cos2u+q^{4m})(1-q^{2m})\nn\\
=2q^{1/4}\cdot\sin u \cdot \prod_{m=1}^{\infty}(1-q^{2m}\cdot e^{2iu})(1-q^{2m}\cdot e^{-2iu})(1-q^{2m})\label{eqC.1}
\eea
\bea
\vartheta_{3}(u)=\prod^{\infty}_{m=1}(1+2q^{2m-1}\cdot\cos2u+q^{2(2m-1)})(1-q^{2m})\nn\\
=\prod^{\infty}_{m=1}(1+q^{2m-1}\cdot e^{2iu})(1+q^{2m-1}\cdot e^{-2iu})(1-q^{2m})\label{eqC.2}
\eea
In particular,
\beq
\vartheta_{3}(0)=\prod_{m=1}^{\infty}(1+q^{2m-1})^{2}(1-q^{2m})\label{eqC.3}
\eeq
For the ratio $\vartheta_{1}(u)/\vartheta_{3}(0)$ we obtain:
\beq
\frac{\vartheta_{1}(u)}{\vartheta_{3}(0)}
=2q^{1/4}\cdot\sin u
\cdot\prod^{\infty}_{m=1}\frac{(1-q^{2m}e^{2iu})(1-q^{2m}e^{-2iu})}{(1+q^{2m-1})^{2}}\label{eqC.4}
\eeq
Taking $\log$ of (\ref{eqC.4}) we get:
\bea
\log\frac{\vartheta_{1}(u)}{\vartheta_{3}(0)}=\frac{1}{4}\log q+\log(i(e^{-iu}-e^{iu}))\nn\\
+\sum_{m=1}^{\infty}[\log(1-q^{2m}e^{2iu})+\log(1-q^{2m}e^{-2iu})-2\log(1+q^{2m-1})]\nn\\
=\frac{1}{4}\log q+i\frac{\pi}{2}-iu+\log(1-e^{2iu})\nn\\
+\sum^{\infty}_{m=1}\sum_{n=1}^{\infty}\frac{(-1)}{n}q^{2mn}\cdot e^{2inu}+\sum_{m=1}^{\infty}\sum_{n=1}^{\infty}\frac{(-1)}{n}q^{2mn}\cdot e^{-2inu}\nn\\
-2\sum_{m=1}^{\infty}\sum_{n=1}^{\infty}\frac{(-1)^{n-1}}{n}q^{2mn-n}\nn\\
=\frac{1}{4}\log q+i(\frac{\pi}{2}-u)+\sum^{\infty}_{n=1}\frac{(-1)}{n}e^{2inu}\nn\\
-\sum^{\infty}_{n=1}\frac{1}{n}\{\frac{q^{2n}\cdot e^{2inu}}{1-q^{2n}}+\frac{q^{2n}\cdot e^{-2inu}}{1-q^{2n}}-2(-1)^{n}\frac{q^{2n}\cdot q^{-n}}{1-q^{2n}}\}\nn\\
=\frac{1}{4}\log q+i(\frac{\pi}{2}-u)\nn\\
-\sum^{\infty}_{n=1}\frac{1}{n}\{\frac{e^{2inu}\cdot(1-q^{2n})+q^{2n}\cdot e^{2inu}+q^{2n}\cdot e^{-2inu}-2(-1)^{n}q^{n}}{1-q^{2n}}\}\label{eqC.5}
\eea
Finally, we obtain:
\bea
\log\frac{\vartheta_{1}(u)}{\vartheta_{3}(0)}=\frac{1}{4}\log q+i(\frac{\pi}{2}-u)\nn\\
-\sum^{\infty}_{n=1}\frac{1}{n}\cdot\frac{e^{2inu}
+q^{2n}\cdot e^{-2inu}-2(-1)^{n}q^{n}}{(1-q^{2n})}\label{eqC>6}
\eea
This is the relation in (\ref{eq5.11}).

\end{document}